\documentclass[12pt,leqno]{article}
\usepackage{amsmath,amsthm}

\def\init{\setcounter{equation}{0}}
\newtheorem{theorem}{Theorem}[section]

\newcommand{\R}{{\bf R}}
\newcommand{\C}{{\bf C}}
\newcommand{\Z}{{\bf Z}}

\newtheorem{lemma}{Lemma}[section]

\newcommand{\e}{{\varepsilon}}

\title{Optical Aharonov-Bohm effect:
an inverse hyperbolic problems approach.
\author{G.Eskin, \ \ \  Department of Mathematics, UCLA,\\ Los Angeles,
CA 90095-1555, USA. \ E-mail: eskin@math.ucla.edu}
}

\begin{document}

\maketitle
\begin{abstract}
We describe the general setting for the optical Aharonov-Bohm effect based on 
the inverse problem of the identification of the coefficients of the governing
hyperbolic equation by the boundary measrements.
We interpret the inverse problem  result as a possibility  in principle to detect
the optical Aharonov-Bohm effect by the boundary measurements.
\end{abstract}

\section{Introduction.}
\label{section 1}
\init
 
In this section we will review the quantum mechanical Aharonov-Bohm (AB) effect
(c.f. [AB],  
[WY], [OP], [W], [E4]).

Let $\Omega$  be a smooth bounded domain in $\R^n$  having the form
$\Omega=\Omega_0\setminus\cup_{j=1}^m\overline{\Omega}_j$,
where $\Omega_0$ is a simply-connected domain and $\Omega_j,1\leq j\leq m,$
are smooth domain called obstacles.  We assume that $\overline{\Omega}_j\subset
\Omega_0$  for $1\leq j\leq m$,  and  
$\overline{\Omega}_j\cap\overline{\Omega}_k=\emptyset$  when $j\neq k,\ 1\leq j,k\leq m$.

Consider the stationary Schr\"{o}dinger equation in $\Omega$   with magnetic potential
$A(x)=(A_1(x),...,A_n(x))$  and  electric potential $V(x)$:
\begin{equation}                                 \label{eq:1.1}
Hu\stackrel{def}{=}\sum_{j=1}^n\left(-i\frac{\partial}{\partial x_j}-A_j(x)
\right)^2u(x)
+V(x)u(x)=k^2u(x),
\end{equation}
describing the nonrelativistic quantum electron in the classical electromagnetic field.
We assume that
\begin{equation}                                \label{eq:1.2}
u|_{\partial\Omega_j}=0,\ \ 1\leq j\leq m,
\end{equation}
i.e. $\Omega_j$  are unpenetrable for the electron, and
\begin{equation}                         \label{eq:1.3}
u|_{\partial\Omega_0}=f(x').
\end{equation}
Let $\Lambda(k) f$  be the Dirichlet-to-Neumann (DN) operator on $\partial\Omega_0$,
i.e.
\begin{equation}                       \label{eq:1.4}
\Lambda(k) f=\left(
\frac{\partial u}{\partial \nu}-i(A\cdot\nu)u
\right)|_{\partial\Omega_0},        
\end{equation}
where $u(x)$  is the solution of (\ref{eq:1.1}), (\ref{eq:1.2}),
(\ref{eq:1.3})
and $\nu$  is the unit outward normal vector  at $x\in \partial\Omega_0$.

Denote  by $G(\overline{\Omega})$  the group of all complex-valued 
$C^\infty(\overline{\Omega})$  functions $c(x)$  in $\overline{\Omega}$
such that $|c(x)|=1$.  

If $c(x)\in G(\overline{\Omega})$  and $u'=c^{-1}(x)u(x)$  then $u'$
satisfies the Schr\"{o}dinger equation of the forn (\ref{eq:1.1})
with $A(x),V(x)$  replaced by $A'(x),V'(x)$,  where
\begin{eqnarray}                          \label{eq:1.5}
A_j'(x)=A_j(x)-ic^{-1}(x)\frac{\partial c}{\partial x_j},\ \ 1\leq j\leq n,
\\
V'(x)=V(x).\ \ \ \ \ \ \ \ \ \ \ \ \ \ \ \ \ \ \ \ \ \ \ \ \ \ \ \ \ \ 
\nonumber
\end{eqnarray}
We shall call the electromagnetic potentials  $A'(x),V'(x)$  and 
$A(x),V(x)$  gauge equivalent.  We also call the DN operators $\Lambda(k)$
and $\Lambda'(k)$,  corresponding to $A(x),V(x)$  and $A'(x),V'(x)$,  respectively, 
gauge equivalent if there exists $c(x)\in G(\overline{\Omega})$  such that
$$
\Lambda'(k)=c_0^{-1}\Lambda(k) c_0,
$$
where $c_0$  is the restriction of $c(x)$  to $\partial\Omega_0$.

Let $B(x)=\mbox{curl\ } A(x)$  or, equivalently,
$\mathcal{B}=d\mathcal{A}$,  where  $\mathcal{A}=\sum_{j=1}^nA_j(x)dx_j$, 
be the magnetic field in $\Omega$.  It follows from (\ref{eq:1.5})
that
$$
B(x)=B'(x)\ \ \mbox{in\ } \ \overline{\Omega}
$$
if $A(x)$  and $A'(x)$  are gauge equivalent.  If   $\Omega$ is
simply-connected then the inverse is true: 
$B(x)=B'(x)$  in $\Omega$  implies  that $A(x)$  and $A'(x)$ are gauge equivalent.
When $\Omega$  is not simply-connected this is not true anymore.
It was shown in the seminal paper of Aharonov and Bohm [AB]  that if
$\mbox{curl\ }A=\mbox{curl\ }A'=0$,  but $A'(x)$  and $A(x)$ belong
to distinct gauge equivalent classes,  they have a different physical impact 
that is detectable in the experiments.
This fact is called the Aharonov-Bohm effect.

An important description of gauge equivalence classes was given by 
Wu and Yang [WY]:

Let $\gamma$  be any closed path in $\overline{\Omega}$.
It is easy to see that $A(x)$  and $A'(x)$  belong to the same gauge equivalent
class iff  
\begin{equation}                       \label{eq:1.6}
\exp(i\int_\gamma A\cdot dx)=\exp(i\int_\gamma  A'\cdot dx)
\end{equation}
for all paths $\gamma$  in $\overline{\Omega}$,
or,  equivalently,
\begin{equation}                       \label{eq:1.7}
\int_\gamma A\cdot dx -\int_{\gamma} A'\cdot dx=2\pi p,
\end{equation}
where $p\in \Z$.

In the original paper [AB] Aharonov and Bahm consider the case of one
obstacle $\Omega_1$  in $\R^2$  and the magnetic field confined to $\Omega_1$.  
Then $\int_\gamma A\cdot dx=\alpha$  is the magnetic flux and $\alpha$
is independent of any simple path $\gamma$ encircling $\Omega_1$.  The quantity 
$e^{i\alpha}$  
that determines the gauge equivalence class of $A(x)$
was measured in this experiment.
If $\alpha\neq 2\pi p,p\in \Z$,   then the gauge equivalence class of $A(x)$  is
nonzero despite 
the fact that $B=0$ in $\Omega=\Omega_0\setminus\overline{\Omega}_1$.  

Consider now the case of several obstacles $\Omega_1,...,\Omega_m$.
Suppose that the  magnetic field is hidden inside  each of these obstacles.
Let $\alpha_k=\int_{\gamma_k}A\cdot dx$  be the magnetic fluxes,
where $\gamma_k$ encircles $\Omega_k$ only.   Suppose that some of
$\frac{\alpha_k}{2\pi}$ are not integers and $\sum_{k=1}^m\alpha_k=0$,
i.e.  the total magnetic flux is zero.  In this case 
the gauge equivalence classes are determined by $m$ parameters $e^{i\alpha_k},1\leq k
\leq m,$  however
the AB experiment will
not find a gauge equivalent class different from zero.
To identify an arbitrary gauge equivalence class one needs to use
broken rays (i.e. the rays reflected at the obstacles)  belonging to the base
of the homotopy group of $\Omega$  (c.f. [E5],  page 1512).

It is necessary to perform at least $m$ AB type experiments
to determine all $e^{i\alpha_k}, 1\leq k\leq m$. When $B(x)=\mbox{curl\ } A$ is
not zero in $\Omega$  it is not enough to perform a finite number of AB type
experiments to identify the gauge equivalence class of $A$.  
Therefore the following question arises:  Is it possible by  the measurements
on the boundary $\partial\Omega_0$ to
 detect the difference in the gauge equivalence classes
of $A(x)$  and $A'(x)$?     The answer to this question is affirmative, and
it is given by the following theorem (c.f. [E4], [W], [N], [KL]  and further
references there):
\begin{theorem}                                        \label{theo:1.1}
Consider two boundary value problems (\ref{eq:1.1}), (\ref{eq:1.2}),
(\ref{eq:1.3})  
corresponding to electromagnetic potentials $A(x),V(x)$  and $A'(x),V'(x)$.
Then $A(x),V(x)$  and $A'(x),V'(x)$
belong to the same gauge equivalence class iff  the DN  operators $\Lambda(k)$  
and
$\Lambda'(k)$  are gauge equivalent for all $k$.
\end{theorem}

We consider each boundary measurement as an experiment.  The Theorem \ref{theo:1.1}
asserts  that the boundary measurements are able to identify an arbitrary gauge 
equivalence class.  We interpret this theorem as a confirmation of the 
Aharonov-Bohm effect.
\qed  

In  \S 2 we develop the same approach  in the case of the optical Aharonov-Bohm 
effect, and we shall  formulate the unique identification theorem  for the optical
AB effect.  In \S 3 
we prove the main unique identification theorem (Theorem  \ref{theo:2.3}).  
Our approach to the hyperbolic 
inverse problems is based on a modification of the BC-method  
given in [E1], [E2].
The powerful BC-method was discovered by M.Belishev and extended by
M.Belishev, Y.Kurylev, M.Lassas and others (c.f. [B], [KKL],   [KL] 
 and additional references there).
An important part of the BC-method is the unique continuation theorem by Tataru[T].
The  approach of [E1], [E2]  allows one to consider 
new problems 
that were not accessible by the BC-method as the inverse
hyperbolic problems with
time dependent coefficients (see [E3]).  The inverse problem results of
this paper are also new.

\section{The optical Aharonov-Bohm effect.}
\label{section 2}
\init

In this section we consider hyperbolic (wave) equation of the form:
\begin{equation}                              \label{eq:2.1}
\sum_{j,k=0}^n\frac{1}{\sqrt{|g|}}\frac{\partial}{\partial x_j}
\left(\sqrt{|g|}g^{jk}(x)\frac{\partial u(x_0,x)}{\partial x_k}\right)=0,
\end{equation}
where $x=(x_1,...,x_n)\in \overline{\Omega},\ x_0$  is the time variable,
$([g^{jk}]_{jk=0}^n)^{-1}$  is the pseudo-Riemannian metric tensor with
Minkowsky signature,  i.e.  the quadratic form
$\sum_{j,k=0}^ng^{jk}(x)\xi_j\xi_k$  has the signature 
$(1,-1,...,-1),\ g(x)=(\det[g^{jk}])^{-1}$.   We assume that $g^{jk}(x)$  are smooth
in $\overline{\Omega}$
and independent of $x_0$.

We make  two additional assumptions:
\begin{equation}                            \label{eq:2.2}
(1,0,...,0) \ \mbox{is a time-like direction, i.e.\ } g^{00}(x)>0,\ 
x \in \overline{\Omega},
\end{equation}
and
\begin{eqnarray}                            \label{eq:2.3}
\ \ \ \ \ \ \ \ \ \ \
\\ 
\mbox{The plane\ \ }\xi_0=0 \mbox{\ intersects the cone\ \ }
\nonumber
\sum_{j,k=0}^ng^{jk}(x)\xi_j\xi_k=0\ \ 
\mbox{at\ \ }  
\\
(\xi_1,\xi_2,...,\xi_n)=(0,0,...,0) \mbox{\ only,}
\nonumber
\\
\mbox{i.e.  the form\ \ }
-\sum_{j,k=1}^n g^{jk}(x)\xi_j\xi_k
\mbox{\ is positive definite,\ \ }
x\in \overline{\Omega}.
\nonumber
\end{eqnarray}
The important physical example of equation of form (\ref{eq:2.1})
 is the equation of the propagation of light in 
the moving medium.  Here the tensor  $g^{jk}(x)$ has the following
form (see Gordon (1923), [NVV], [LP1]):
\begin{equation}                        \label{eq:2.4}
g^{jk}=\eta^{jk}+(n^2(x)-1)u^ju^k,\ \ 0\leq j,\ k\leq n, \ n=3,
\end{equation}
when $[\eta^{jk}]^{-1}$ is the Lorentz metric tensor,  $\eta^{jk}=0$,
when $j\neq k,\ \eta^{00}=1,\ \eta^{jj}=-1$ for $1\leq j\leq n,
\ x_0=ct,\ n(x)=\sqrt{\e(x)\mu(x)}$  is the refraction index,
$(u^0,u^1,u^2,u^3)$  is the four-velocity of the medium flow, 
$(u^0,u^1,u^2,u^3)= 
(1-\frac{|w|^2}{c^2})^{-\frac{1}{2}}(1,\frac{w}{c}),
\ w(x)=(w_1,w_2,w_3)$ is the  velocity of the
flow (c.f. [LP], [LP1], [LP2]).

In the case of slowly moving medium one drops the terms of order $(\frac{|w|}{c})^2$
(c.f. [LP1], [LP2], [CFM]).  Then the metric of the slowly moving medium has the form:
\begin{eqnarray}                               \label{eq:2.5}
g^{jk}=\eta^{jk}\ \ \ \mbox{for\ }\ \ \ \ 1\leq j,k\leq n,
\\
g^{00}=n^2(x),\ \ \ g^{0j}=g^{j0}=v_j(x)
\stackrel{def}{=}(n^2-1)\frac{w_j(x)}{c},\  
\ 1\leq j\leq n, n=3,
\nonumber
\end{eqnarray}
and the corresponding equation is
\begin{eqnarray}                              \label{eq:2.6}
\ \ \ \ \ \ \ \ \ \ \ \ \ \ \ \ \ \ \ \  
n^2(x)\frac{\partial^2 u}{\partial x_0^2} +\sum_{j=1}^n\frac{1}{\sqrt{|g(x)|}}
\frac{\partial}{\partial x_j}\left(\sqrt{|g(x)|}v_j(x)\frac{\partial u}{\partial x_0}\right)
\ \ \ \ \ \ \ \ \ \ \ \ \ \ \ \ \ 
\\
+\sum_{j=1}^n\frac{1}{\sqrt{|g(x)|}}
\frac{\partial}{\partial x_0}\left(\sqrt{|g(x)|}v_j(x)\frac{\partial u}{\partial x_j}\right)
\nonumber
-\sum_{j=1}^n\frac{1}{\sqrt{|g(x)|}}
\frac{\partial}{\partial x_j}\left(\sqrt{|g(x)|}\frac{\partial u}{\partial x_j}\right)
=0.
\end{eqnarray}

We shall also consider in addition to the equation (\ref{eq:2.6})
 the following equation: 
\begin{equation}                              \label{eq:2.7}
n^2(x)\frac{\partial^2 u}{\partial x_0^2} -\sum_{j=1}^n\frac{1}{\sqrt{|g(x)|}}
\left(\frac{\partial}{\partial x_j}-v_j(x)\frac{\partial}{\partial x_0}\right)
\sqrt{|g(x)|} 
\left(\frac{\partial}{\partial x_j}-v_j(x)\frac{\partial}{\partial x_0}\right)u
=0,
\end{equation}
where
$v_j(x)$  and $n^2(x)$  are the same as in (\ref{eq:2.5}),  $1\leq j\leq n$.

Equation (\ref{eq:2.7})  differs from the equation  (\ref{eq:2.6})
 by the term 
$\sum_{j=1}^nv_j^2(x)\frac{\partial^2 u}{\partial x_0^2}$.  Since 
$v_j^2=O((\frac{|w|}{c})^2)$  the equation (\ref{eq:2.7}) also
describes the propagation of light in the slowly moving medium.
We consider (\ref{eq:2.7})   to have a closer 
analogy with the quantum mechanical AB effect,  although the addition
of extra terms affects the uniqueness of the inverse problem
(compare Theorems \ref{theo:2.1} and \ref{theo:2.2}).
Note that the nonuniqueness is of the first order in $\frac{|w|}{c}$
(see Theorem \ref{theo:2.1}). 

We consider the initial-boundary value problem for 
(\ref{eq:2.6})  and (\ref{eq:2.7}) 
in the infinite cylinder $\Omega\times(-\infty,+\infty)$,
where  $\Omega$  is the same domain as in \S 1:
\begin{equation}                              \label{eq:2.8}
u(x_0,x)=0\ \ \ \mbox{for\ \ } x_0 \ll 0,
\end{equation}
\begin{eqnarray}                             \label{eq:2.9}
u(x_0,x)|_{\partial \Omega_j\times(-\infty,+\infty)}=0,
\ \ 1\leq j\leq m,
\\
\nonumber
u(x_0,x)|_{\partial \Omega_0\times(-\infty,+\infty)}=f(x_0,x'),\ \ 
x'\in\partial\Omega_0,
\end{eqnarray}
where $f(x_0,x')$   has a compact support on 
$\partial\Omega_0\times(-\infty,+\infty)$.

Denote by $\Lambda$  the hyperbolic
DN operator:
\begin{equation}                        \label{eq:2.10}
\Lambda f=\left(\frac{\partial u}{\partial \nu}-
(v\cdot\nu)\frac{\partial u}{\partial x_0}
\right)|_{\partial \Omega_0\times(-\infty,+\infty)},
\end{equation}
where,  as in \S1,  $\nu$  is the external unit normal to  
$\partial\Omega_0$.  

In studying the 
equation (\ref{eq:2.7})
we shall use the following change of variables in $\Omega\times(-\infty,+\infty)$:
\begin{equation}                                \label{eq:2.11}
\hat{x}_0=x_0+a(x),\ \ \ \ \hat{x}_j=x_j,\ 1\leq j\leq n,
\end{equation}
where $a(x)\in C^\infty(\overline{\Omega}),\ a(x)=0$ on $\partial\Omega_0$.
If $\hat{u}(\hat{x}_0,\hat{x})$  is $u(x_0,x)$  in new coordinates,
then $\hat{u}(\hat{x}_0,\hat{x})$ also satisfies an equation of the form
(\ref{eq:2.7}):
\begin{eqnarray}                  \label{eq:2.12}
\hat{L}\hat{u}\stackrel{def}{=}                              
\hat{n}^2(x)\frac{\partial^2 \hat{u}(\hat{x}_0,x)}{\partial \hat{x}_0^2} 
\ \ \ \ \ \ \ \ \ \ \ \ \ \ \ \ \ \ \ \ \ \ \ \ \ \ \ 
\ \ \ \ \ \ \ \ \ \ \ \ \ \ \ \ \ \ \ 
\\
-\sum_{j=1}^n\frac{1}{\sqrt{|\hat{g}(x)|}}
\left(\frac{\partial}{\partial x_j}-\hat{v}_j(x)\frac{\partial}{\partial \hat{x}_0}\right)
\sqrt{|\hat{g}(x)|} 
\left(\frac{\partial}{\partial x_j}
-\hat{v}_j(x)\frac{\partial}{\partial \hat{x}_0}\right)
\hat{u}=0,
\nonumber
\end{eqnarray}
where $\hat{n}(x)=n(x)$,  
$\ v_j(x)$ is replaced by
\begin{equation}                         \label{eq:2.13}
\hat{v}_j(x)=v_j(x)-a_{x_j}(x),\ 1\leq j\leq n.
\end{equation}
We assume that $\sum_{j=1}^n\hat{v}_j^2(x)<n^2(x)$  to preserve the
hyperbolicity of (\ref{eq:2.12}).
  Note that
\begin{equation}                    \label{eq:2.14}
\hat{u}=0 \ \ \mbox{for\ \ \ } \hat{x}_0\ll 0
\end{equation}
and 
\begin{eqnarray}                             \label{eq:2.15}
\hat{u}|_{\partial \Omega_j\times(-\infty,+\infty)}=0,
\ \ 1\leq j\leq m,
\\
\hat{u}|_{\partial \Omega_0\times(-\infty,+\infty)}=\hat{f}(\hat{x}_0,x'), 
\nonumber
\end{eqnarray}
where $\hat{f}(\hat{x}_0,x')=f(x_0,x')$   since $a=0$  on
$\partial\Omega_0$

We shall say that $\hat{v}_j,\ 1\leq j\leq n$,  and $v_j,\ 1\leq j\leq n$,  
belong to the same equivalence class if (\ref{eq:2.13}) holds.

If $v(x)=(v_1(x),...,v_n(x))$  and $\hat{v}(x)=(\hat{v}_1,...,\hat{v}_n)$
belong to the same equivalence class then 
\begin{equation}                           \label{eq:2.16}
\int_\gamma v\cdot dx-\int_\gamma\hat{v}\cdot dx =0
\end{equation}
for all closed paths in $\overline{\Omega}$   since 
$\sum_{j=1}^n\int_\gamma a_{x_j}dx_j=0$. 

It is easy to see that if $v$ and $\hat{v}$  belong to the same equivalence
class then the $DN$  operators $\Lambda$  and $\hat{\Lambda}$ are
equal on $\partial\Omega_0\times(-\infty,+\infty)$.  A nontrivial fact is
that the inverse
statement
 is also true.  The following unique identification theorem holds:

\begin{theorem}                                \label{theo:2.1}
Let $Lu=0,\ \hat{L}\hat{u}=0$  be equations of the form (\ref{eq:2.7}),
(\ref{eq:2.12})  in domains $\Omega,\hat{\Omega}=\Omega_0\setminus\cup_{j=1}^{\hat{m}}
\hat{\Omega}_j$,  respectively.  Suppose that
 the DN operators $\Lambda$  and $\hat{\Lambda}$  are equal 
on $\partial\Omega_0\times(-\infty,+\infty)$  for all
$f\in C_0^\infty(\partial\Omega_0\times(-\infty,+\infty))$.
Then the $\hat{\Omega}=\Omega,\ \hat{n}(x)=n(x)$ and the 
corresponding velocity flows $v(x),\hat{v}(x)$
belong to the same equivalent class,  i.e. 
(\ref{eq:2.13}) holds for some $a(x)\in C^\infty(\overline{\Omega}),
\ a(x)|_{\partial \Omega_0 }=0$.
\end{theorem}
Note that we did not assume apriori that the number of obstacles
$\hat{m}$ in $\hat{\Omega}$ and their location are  the same as in $\Omega$.   

A consequence of Theorem \ref{theo:2.1}  is that  boundary measurements on 
$\partial\Omega_0\times(-\infty,+\infty)$  uniquely determine the integrals
$\int_\gamma v\cdot dx$  for all  paths $\gamma$  in $\overline{\Omega}$.

As in \S 1 we view the  optical Aharonov-Bohm effect as the fact
that the different equivalence classes of the velocity flow have different
physical impacts. Theorem \ref{theo:2.1} confirms that the boundary measurement
(experiments)  allow one to distinguish different equivalence classes,
i.e.  to detect the  Aharonov-Bohm effect.

{\bf Remark 2.1}
There is a  difference between the optical Aharonov-Bohm effect
and the quantum mechanical AB effect.   In the case of the optical AB effect
the boundary measurements allow  one to recover  $\int_\gamma v\cdot dx$.  In
the  case of the quantum mechanical AB effect we can recover only 
$\int_\gamma A\cdot ~dx \ \ 
\mbox{(mod $\ 2\pi p$)},\\ p\in\Z$.
\qed

Let $\tilde{u}(k,x))=\int_{-\infty}^\infty u(x_0,x)e^{-ix_0k}dx_0$  
be the Fourier-Laplace transform of $u(x_0,x)$ in $x_0$,
or let  $e^{ikx_0}\tilde{u}(k,x)$  be a monochromatic wave.
Then $\tilde{u}(k,x)$  satisfies  the Schr\"{o}dinger equation:
\begin{eqnarray}                              \label{eq:2.17}
-k^2n^2(x)\tilde{u}(k,x) 
-\sum_{j=1}^n\frac{1}{\sqrt{|g(x)|}}
\left(\frac{\partial}{\partial x_j}-ikv_j(x)
\right)
\\
\cdot\sqrt{|g(x)|} 
\left(\frac{\partial}{\partial x_j}-ikv_j(x)
\right)\tilde{u}(k,x)=0
\nonumber
\end{eqnarray}
with the boundary conditions
\begin{eqnarray}                       \label{eq:2.18}
\tilde{u}(k,x)|_{\partial\Omega_j}=0,
\\
\tilde{u}(k,x)|_{\partial\Omega_0}=\tilde{f}(k,x').
\nonumber
\end{eqnarray}
Now $kv(x)$  plays the role of the vector potential and it depends  on $k$.
Note also that the 
Fourier-Laplace image $\tilde{T}$ of the  transformation  (\ref{eq:2.11})  is
the multiplication  by $e^{ika(x)}$,
i.e. $\tilde{T}$  is a gauge transformation depending on parameter $k$.

When $\Omega$ is multi-connected one can expect that the Aharonov-Bohm effect
takes place for (\ref{eq:2.17}).
This problem was studied in optics (c.f. [LP], [LP1], [LP2], [CFM]).
An analogous problem was considered for the water waves and for 
the acoustic
waves in [BCLUW], [RdeRTF], [VMCL].

These authors considered the case of one obstacle $\Omega_1\subset\R^2$
and  irrotational flow  in
$\Omega_0\setminus\Omega_1$.
Performing an Aharonov-Bohm type experiment they measured 
$\exp({i \int_{\gamma} v\cdot dx})$  
as in the quantum mechanical  AB effect.
Since such experiments are based on the geometric optics considerations
it was assumed that the light rays are straight lines and $kv(x)$  is not 
large.
A natural question arises whether some form  of the AB effect
takes place when these conditions are not satisfied.

Note that
a rigorous
 geometric optics approach when $k\rightarrow\infty$ for the equation 
(\ref{eq:2.17}) is  more 
delicate than for the equation (\ref{eq:1.1}).  
In particular,  the eiconal equation
depends on $v(x)$  and the light rays  are not the straight lines.

{\bf Remark 2.2}
Let $\mbox{curl\ } v=0$  in $\Omega=(\Omega_0\setminus\overline{\Omega_1}))
\subset\R^2$.
In this case the equivalence class of $v(x)$ depends only on one 
parameter $\alpha=\int_\gamma v\cdot dx$,  where $\gamma$ encircles
$\Omega_1$.
There is  a simple solution of the inverse problem in this case that
does not use neither the geometric optics  nor the Theorem \ref{theo:2.1}:

Let $v(x)$ and $\hat{v}(x)$  be two irrotational velocity flows in
$\Omega_0\setminus\overline{\Omega}_1$.
Consider two Schr\"{o}dinger equations  of the form (\ref{eq:2.17})  in
$\Omega=\Omega_0\setminus\overline{\Omega}_1$  assuming that $\hat{n}(x)=n(x)$
in $\overline{\Omega}$  
and $\Lambda(k)=\hat{\Lambda}(k)$  on $\partial\Omega_0$  for some fixed $k$.
It was shown in [NSU], [ER],  using the parametrix of the DN operators,
that 
$\hat{v}\cdot\tau=v\cdot \tau$  on $\partial\Omega_0$,
where  $\tau(x)$ is  the tangent vector to
$\partial\Omega_0$ at $x\in\partial\Omega_0$.
It follows from $\hat{v}\cdot \tau= v\cdot \tau$  on $\partial\Omega_0$
that $\alpha=\int_{\partial\Omega_0}v\cdot dx=\int_{\partial\Omega_0}\hat{v}\cdot dx$.
Since
$v$  and $\hat{v}$  are irrotational this implies that there exists $a(x)\in C^\infty(\overline{\Omega})$ such that $\hat{v}-v=\frac{\partial a}{\partial x}$.
Since $\frac{\partial a}{\partial x}\cdot\tau=0$  on $\partial\Omega_0$
we get that $a|_{\partial\Omega_0}=a_0=\mbox{const}$.
Replacing $a(x)$  by $a(x)-a_0$  we obtain that $\hat{v}$   and $v$
belong to the same equivalence class.

Similar arguments apply in the case of equations (\ref{eq:2.6})  and
(\ref{eq:2.1})  with the metric (\ref{eq:2.4}).   Using the parametrix of
the DN  operator we can recover the restriction of the metric to 
$\partial\Omega_0$  (c.f.  [LU] or [E1],  Remark 2.2).  In particular,
we can determine $w(x)\cdot\tau(x)$  on $\partial\Omega_0$.
Therefore we can recover $\alpha=\int_{\partial\Omega_0} w(x)\cdot dx$.  
In the case of irrotational flow and one obstacle $\alpha$  is the same
for any simple path in $\Omega=\Omega_0\setminus\overline{\Omega}_1$.
\qed

We shall investigate now the inverse  problem for the equation (\ref{eq:2.6}).
The case of the equation (\ref{eq:2.1}) 
with the metric (\ref{eq:2.4}) will be studied in another paper.

\begin{theorem}                      \label{theo:2.2}
Consider two initial-boundary value problems in domains 
$\Omega\times(-\infty,+\infty)$ and $\hat{\Omega}\times(-\infty,+\infty)$
for operators of the form (\ref{eq:2.6}), corresponding to the  metric
tensors $[g^{jk}(x)]^{-1},\ [\hat{g}^{jk}(\hat{x})]^{-1}$  
of the form (\ref{eq:2.5}),
respectively.
Assume that the DN operators $\Lambda$  and $\hat{\Lambda}$,  corresponding to
$L$  and $\hat{L}$ are equal on $\partial\Omega_0\times
(-\infty,+\infty)$.
Assume also that 
there exists an open dense set $O\subset\Omega$  such that
the velocity flow $\hat{v}(x)=(\hat{v}_1,...,\hat{v}_n)$
does not vanish on $O$.  Then
$$
\hat{\Omega}=\Omega,\ \ \hat{n}(x)=n(x),\ \ \hat{v}(x)=v(x),\ \ \ 
1\leq j\leq n,
$$
unless $\hat{v}(x)$  is a gradient flow,  i.e. 
there exists $b(x)\in C^\infty(\overline{\Omega})$ such that 
$\hat{v}(x)=\frac{\partial b}{\partial x}$  and $b(x)=0$  on 
$\partial\Omega_0$.
In the case of the gradient flow there are two solutions $\hat{v}(x)=v(x)$
and $\hat{v}(x)=-v(x)$.
\end{theorem}
The proofs of Theorems \ref{theo:2.1}  and \ref{theo:2.2}  will be given
in the end of this section.
\qed

Now we shall consider the general case of the initial-boundary value
problem (\ref{eq:2.8}), (\ref{eq:2.9})  for the equation (\ref{eq:2.1}).
The DN operator for (\ref{eq:2.1}) has the following form:
\begin{equation}                                      \label{eq:2.19}
\Lambda f=\sum_{j,k=0}^n g^{jk}(x)\frac{\partial u}{\partial x_j}\nu_k
\left|\sum_{p,r=1}^ng^{pr}(x)
\nu_p\nu_r\right|^{-\frac{1}{2}}|_{\partial\Omega_0\times(-\infty,+\infty)},
\end{equation}
where
$\nu$  is the unit normal as in (\ref{eq:2.10}).

Consider the diffeomorphism of the form:
\begin{eqnarray}                          \label{eq:2.20}
\hat{x}_0=x_0+a(x),
\\
\nonumber
\hat{x}=\varphi(x),
\end{eqnarray}
  where $a(x)|_{\partial\Omega_0}=0$  and
$\varphi(x)$  is a diffeomorphism of $\overline{\Omega}$  onto 
$\overline{\hat{\Omega}}$, 
where $\hat{\Omega}$ is a domain of the form 
$\hat{\Omega}=\Omega_0\setminus\cup_{j=1}^{\hat{m}}\overline{\hat{\Omega}}_j$
  and $\varphi=I$
on $\partial\Omega_0$.
Note that (\ref{eq:2.20})  transforms (\ref{eq:2.1})  into the equation
of the same form.  More precisely,  (\ref{eq:2.1})  has the following form
in $(\hat{x}_0,\hat{x})$  coordinates:
\begin{equation}                           \label{eq:2.21}
\hat{L}\hat{u}=\sum_{j,k=0}^n\frac{1}{\sqrt{|\hat{g}(\hat{x})|}}
\frac{\partial}{\partial \hat{x}_j}\left(\sqrt{|\hat{g}(\hat{x})|}
\hat{g}^{jk}(\hat{x})
\frac{\partial \hat{u}}{\partial \hat{x}_k}\right)=0,
\end{equation}
where 
\begin{eqnarray}                         \label{eq:2.22}
[\hat{g}^{jk}(\hat{x})]=
J(x)[g^{jk}(x)]J^T(x),
\\
\nonumber
\hat{g}(\hat{x}) =(\det[\hat{g}^{jk}(\hat{x})])^{-1},
\ \ 0\leq j,k\leq n,
\
\end{eqnarray}
$J(x)$  is the Jacobi matrix of (\ref{eq:2.20}).

\begin{theorem}                       \label{theo:2.3}
Consider 
equations (\ref{eq:2.1}) and (\ref{eq:2.21}) in domains 
$\Omega\times(-\infty,+\infty)$  and $\hat{\Omega}\times(-\infty,+\infty)$,
respectively,  with initial-boundary conditions
(\ref{eq:2.8}), (\ref{eq:2.9})  and  (\ref{eq:2.14}), (\ref{eq:2.15}),
respectively,  where $f=\hat{f}$.   Assume  that the DN operators $\Lambda$  
and $\hat{\Lambda}$ are equal on 
$\partial\Omega_0\times(-\infty,+\infty)$  and the conditions 
(\ref{eq:2.2}), (\ref{eq:2.3})  hold for $L$ and $\hat{L}$.
Then there exists a map $\psi$  of the form (\ref{eq:2.20})  such that
\begin{equation}                       \label{eq:2.23}
  \psi\circ\hat{L}=L\ \ \ \mbox{in\ \ } \Omega\times(-\infty,+\infty).
\end{equation}
\end{theorem}
Note that (\ref{eq:2.23}) is equivalent to (\ref{eq:2.22}).  Note also
that since $\varphi$  is a diffeomorphism of $\overline{\Omega}$  onto
$\overline{\hat{\Omega}}$,   we have that $\hat{m}=m$  and $\partial\Omega_j$  are
diffeomorphic to $\partial\hat{\Omega}_j,\ 1\leq j\leq m$.

The proof of Theorem \ref{theo:2.3}  will be given in \S 3.  

{\bf Remark 2.3}.
Making   the Fourier-Laplace transform
in (\ref{eq:2.1}) we obtain
\begin{equation}                        \label{eq:2.24}
L(ik,\frac{\partial}{\partial x})\tilde{u}(k,x)=0\ \ \ \mbox{in}\ \ \Omega,
\end{equation}
where $L(\frac{\partial}{\partial x_0},\frac{\partial}{\partial x})$  is
the operator (\ref{eq:2.1}).  Let $\Lambda(k)$ be the Fourier-Laplace image
of the DN operator (\ref{eq:2.19}).   Using well known estimates for the
initial-boundary value problem  (\ref{eq:2.1}), (\ref{eq:2.8}), (\ref{eq:2.9}) one 
can prove that the hyperbolic  DN operator (\ref{eq:2.19}) on  
$\partial\Omega_0\times(-\infty,+\infty)$   uniquely determines the DN operator 
$\Lambda(k)$  for the elliptic boundary value problem (\ref{eq:2.24}), (\ref{eq:2.18})
and vice versa  (see, for example,  [KKLM]).
Here $k\in \C\setminus\Z$,  where $\Z$ is a discrete set.

Suppose $g^{jk}-\eta^{jk}=0$,    when $|x|>R$,   and suppose that
$\Omega_0\supset\{x:|x|\leq R\}$.  It is well known that $\Lambda(k)$
given on $\partial\Omega_0$  for fixed $k=k_0$  uniquely determines 
the scattering amplitude $a(\theta,\omega,k)$  for $k=k_0$  and any
$\theta\in S^{n-1},\ \omega\in S^{n-1}$,  and vice versa  (see,  for example,
the recent work [OD] and additional references there). 

Therefore one can consider the inverse scattering problem
for
(\ref{eq:2.24})  in $\R^n$  instead of the inverse boundary value problem for
(\ref{eq:2.24}),  (\ref{eq:2.18}).   In the case when
  there is no obstacles and the principal part of
(\ref{eq:2.24})  is the Laplacian,  such  inverse problems were studied
for  $n\geq 3$  and fixed $k$
(see,  for example, [NSU] and [ER1],  where the case of exponentially decreasing 
electromagnetic potentials was considered).   When  obstacles are present
or when the metric is not Euclidean the hyperbolic inverse problem approach
is much more powerful.
\qed

We shall show
now how Theorem \ref{theo:2.3}  implies Theorem \ref{theo:2.1}
and Theorem \ref{theo:2.2}.

{\bf Proof of Theorem 2.1}
Consider two equations of the form 
(\ref{eq:2.7})  and (\ref{eq:2.12}),  i.e. $g^{jk}=-\delta_{jk}$  in
(\ref{eq:2.1})  and 
$\hat{g}^{jk}=-\delta_{jk}$  in
(\ref{eq:2.21}),  $1\leq j,k\leq n$.   We  assume   
that $\Lambda=\hat{\Lambda}$  on $\partial\Omega_0\times(-\infty,+\infty)$.
It follows from Theorem \ref{theo:2.3}
that there exists a map $\psi$  of the form (\ref{eq:2.20})   such that 
(\ref{eq:2.22}) holds.   It follows from (\ref{eq:2.22}) that
\begin{equation}                         \label{eq:2.25}
\hat{g}^{jk}=\sum_{p,r=1}^n g^{pr}\frac{\partial\varphi_j}{\partial x_p}
\frac{\partial\varphi_k}{\partial x_r},\ \ \ 
1\leq j,k \leq n,
\end{equation}
where
\begin{equation}                        \label{eq:2.26}
\varphi=(\varphi_1,...,\varphi_n)=I\ \ \mbox{on\ }  \partial\Omega_0.
\end{equation} 
Since 
$\hat{g}^{jk}=-\delta_{jk},\ g^{pr}=-\delta_{pr}$  we have that
$\varphi_j(x)=x_j,\ 1\leq j\leq n,$  is the solution of
(\ref{eq:2.25})  and the uniqueness of the Cauchy problem
(\ref{eq:2.25}), (\ref{eq:2.26})implies that $\varphi=I$  is the only
solution  of (\ref{eq:2.25}), (\ref{eq:2.26}).
Therefore the map (\ref{eq:2.20})  reduces to the map (\ref{eq:2.11}).  
This implies  that $\hat{\Omega}=\Omega$.  
Making the change of variables (\ref{eq:2.11}) with the same $a(x)$ as 
in (\ref{eq:2.20}) we  get two identical operators.  Therefore (\ref{eq:2.13})
holds and,  subseqently, $\hat{n}(x)=n(x)$.
\qed

{\bf Proof of Theorem 2.2}.
It follows from Theorem \ref{theo:2.3} that there exists a map of the form 
(\ref{eq:2.20})  such that (\ref{eq:2.22})  holds.  Let
$[g_{jk}(x)]=[g^{jk}(x)]^{-1}$,\\ $[\hat{g}_{jk}(\hat{x})]=
[\hat{g}^{jk}(\hat{x})]^{-1}$.   Then (\ref{eq:2.22})  is equivalent to
\begin{equation}                       \label{eq:2.27}
\sum_{j,k=0}^n\hat{g}_{jk}(\hat{x})d\hat{x}_jd\hat{x}_k=
\sum_{j,k=0}^n g_{jk}(x)dx_jdx_k,
\end{equation}
where $(\hat{x}_0,\hat{x})$  are related to $(x_0,x)$  by (\ref{eq:2.20}).
Note that (c.f. [LP1])
\begin{eqnarray}                         \label{eq:2.28}
g_{00}=n^{-2}(x),\ \ g_{jk}=-\delta_{jk}\ \ \mbox{for\ \ } 1\leq j,k \leq n,
\\
g_{0j}=g_{j0}=-(n^{-2}(x)-1)\frac{w_j(x)}{c}=n^{-2}(x)v_j(x),
\nonumber
\end{eqnarray}
and $\hat{g}_{jk}$  have a similar form. Here $v_j(x)$  is the same as in 
(\ref{eq:2.5}).
Since $g^{jk}=\hat{g}^{jk}=-\delta_{jk}$  for  $1\leq j,k\leq n$,  we have,
as in  the proof of Theorem \ref{theo:2.1},   
that $\hat{x}=\varphi(x)=x$.   Therefore $\hat{\Omega}=\Omega$.  Note that
\begin{equation}                        \label{eq:2.29}
d\hat{x}_0=dx_0+
\sum_{j=1}^n
\frac{\partial a(x)}{\partial x_j}dx_j.
\end{equation}
Substitute (\ref{eq:2.29})  into (\ref{eq:2.27}).  Taking into account that
$\hat{x}_j=x_j,\ 1\leq j\leq n,$  and that $dx_0, dx_1,...,dx_n$  are
arbitrary,  we get  from (\ref{eq:2.27})  and (\ref{eq:2.29}):
\begin{equation}                             \label{eq:2.30}
\hat{n}^{-2}(x)=n^{-2}(x),
\end{equation}
\begin{equation}                             \label{eq:2.31}
2\hat{n}^{-2}(x)a_{x_j}+2\hat{n}^{-2}(x)\hat{v}_j(x)=2n^{-2}(x)v_j(x),
\ \ 1\leq j\leq n,
\end{equation}         
\begin{equation}                             \label{eq:2.32}
\hat{n}^{-2}(x)(\sum_{j=1}^na_{x_j}dx_j)^2+
2(\sum_{j=1}^n\hat{n}^{-2}(x)\hat{v}_j(x)dx_j)(\sum_{j=1}^n a_{x_j}dx_j)=0.
\end{equation}
It follows
from (\ref{eq:2.30})  that $\hat{n}(x)=n(x)$.   Multiplying (\ref{eq:2.31}) by
$n^2(x)$  we get
\begin{equation}                            \label{eq:2.33}
\hat{v}_j(x)+a_{x_j}(x)=v_j(x),\ 1\leq j\leq n.
\end{equation}
If 
there exists
$x$   such that not all $a_{x_j}(x)=0,\ 1\leq j\leq n,$
then we can cancel $\hat{n}^{-2}(x)\sum_{j=1}^na_{x_j}(x)dx_j$  in 
(\ref{eq:2.32}) and get  $\sum_{j=1}^n a_{x_j}(x)dx_j+
2\sum_{j=1}^n\hat{v}_j(x)dx_j=0$,  i.e.
\begin{equation}                           \label{eq:2.34}
a_{x_j}(x)+2\hat{v}_j(x)=0,\ \ 1\leq j\leq n,
\end{equation}
since $dx_j,\ 1\leq j\leq n,$  are arbitrary.  
Comparing (\ref{eq:2.33}) and (\ref{eq:2.34})  we get
$$
\hat{v}_j(x)=-v_j(x),\ \ 1\leq j\leq n,
$$
when $\frac{\partial a}{\partial x}=
(\frac{\partial a}{\partial x_1},...,\frac{\partial a}{\partial x_n})\neq 0$.
 In the case when 
$\hat{v}(x)\neq 0$ on an open dense set in $\Omega$  we have, 
by connectness of $\Omega$ and 
 by the 
continuity,  that (\ref{eq:2.34}) holds on $\overline{\Omega}$
if $a(x)\not\equiv 0$.
Therefore 
$\hat{v}(x)$  is a gradient flow,
$v(x)=-\hat{v}(x)$  is also  a solution of the inverse problem,
and it is the only solution except the trivial solution $v(x)=\hat{v}(x)$
that corresponds to $a(x)=0$.
Note 
that the boundary measurements can not distinguish  between these  two solutions 
$v(x)$  and $-v(x)$.
If $\hat{v}=0$  on an open set in $\Omega$ then (\ref{eq:2.32})
implies that $\frac{\partial a }{\partial x}=0$  on this set.
In such case there can be more than two solutions of the inverse problem.
For example,
 if $v(x)$ is a gradient flow,  i.e $\hat{v}=
\frac{\partial b}{\partial x},\ b(x)=0$  on $\partial\Omega_0$
and the closure of the set $\{x\in\Omega:b(x)\neq 0\}$  is not connected,
then there exists at least four solutions of the inverse problem.  

If $v(x)$  and $\hat{v}(x)$  are any two solutions of the inverse problem
then (\ref{eq:2.33}) implies that $\int_\gamma\hat{v}(x)\cdot dx=
\int_\gamma v(x)\cdot dx $  for any $\gamma$  in $\Omega$.  Therefore
the boundary measurements uniquely determine $\int_\gamma v(x)\cdot dx$.
This fact can be  considered as an analogue
of the Aharonov-Bohm effect.

\section{The proof of the main theorem.}
\label{section 3}
\init
 
As in [E1] we start the proof of
Theorem \ref{theo:2.3} with the introduction of a convenient system of cooordinates 
that simplifies the equation.

Let $U_0$ be a neighborhood of some part $\Gamma$  of $\partial\Omega_0$  and
let
$(x',x_n)$ be a system of coordinates in $U_0$ such that $x_n=0$ is the equation
of $\partial\Omega_0\cap U_0$.
Let $T$ be small.

Denote by $\psi^\pm,$
the solutions of  the eiconal equations in $U_0$
\begin{equation}                                      \label{eq:3.1}
\sum_{j,k=0}^ng^{jk}(x)\psi_{x_j}^\pm(x_0,x)\psi_{x_k}^\pm(x_0,x)=0,
\end{equation}
such that
\begin{eqnarray}                                     \label{eq:3.2}
\psi^+=x_0\ \ \ \ \mbox{when\ }\ \ x_n=0,
\\
\psi^-=T-x_0\ \ \ \ \mbox{when\ }\ \ x_n=0,
\nonumber
\end{eqnarray}

\begin{equation}                                     \label{eq:3.3}
\psi^\pm_{x_n}|_{x_n=0}=
\frac{\mp g^{0n}(x)+\sqrt{(g^{0n}(x))^2-g^{00}(x)g^{nn}(x)}}{g^{nn}(x)}|_{x_n=0},
\end{equation}
Solutions  $\psi^\pm(x_0,x)$ exist for $0\leq x_n\leq \delta$ where  $\delta$
is small.
We assume that surfaces $\psi^+=0$ and $\psi^-=0$  intersect   when $x_n\leq \delta$.

In the case when $g^{jk}(x)$ are independent of $x_0$ we have 
\begin{eqnarray}                                     \label{eq:3.4}
\psi^+=x_0 +\varphi^+(x),
\\
\psi^-=T-x_0  +\varphi^-(x),
\nonumber
\end{eqnarray}
where $\varphi^\pm(x)$  satisfy
\begin{eqnarray}                                     
g^{00}(x)\pm 2\sum_{j=1}^n g^{0j}(x)\varphi_{x_j}^\pm +
\sum_{j,k=1}^ng^{jk}(x)\varphi_{x_j}^\pm\varphi_{x_k}^\pm=0,
\nonumber
\\
\varphi^\pm|_{x_n=0}=0,\ \ \ 
\varphi^\pm_{x_n}|_{x_n=0}=
\frac{\mp g^{0n}(x)+\sqrt{(g^{0n}(x))^2-g^{00}(x)g^{nn}(x)}}{g^{nn}(x)}|_{x_n=0}.
\nonumber
\end{eqnarray}

Denote by $\varphi_p(x)$  the solutions of 
\begin{equation}                              \label{eq:3.5}
\sum_{j,k=0}^ng^{jk}(x)\psi_{x_j}^-\varphi_{px_k}(x)=0
\end{equation}
with the  initial conditions
\begin{equation}                             \label{eq:3.6}
\varphi_p|_{x_n=0}=x_p,\ \ 1\leq p\leq n-1.
\end{equation}

Note that $\varphi_{px_0}=0,\psi_{x_0}^-=-1$.    Therefore we have
$$
\sum_{j,k=1}^ng^{jk}\varphi_{x_j}^-\varphi_{px_k}-\sum_{j=1}^n
 g^{j0}\varphi_{px_j}=0,
\ \ 1\leq p\leq n-1.
$$

Make the following change of variables in 
$U_0\times [0,T]$:
\begin{eqnarray}                         \label{eq:3.7}
s=\psi^+(x_0,x)=x_0+\varphi^+(x),
\\
\tau=\psi^-(x_0,x)=T-x_0+\varphi^-(x),
\nonumber
\\
y_j=\varphi_j(x),\ \ 1\leq j\leq n-1.
\nonumber
\end{eqnarray}
We shall call $(s,\tau,y')$  the Goursat coordinates.

Let 
$
\hat{u}(s,\tau,y')=u(x_0,x).
$
Then $\hat{u}(s,\tau,y')$  satisfies the equation
\begin{eqnarray}                         \label{eq:3.8}
\\
\
\hat{L}\hat{u}\stackrel{def}{=}
-\frac{2}{\sqrt{|\hat{g}|}}
\frac{\partial}{\partial s}\left(\hat{g}^{+,-}(s,\tau,y')
\sqrt{|\hat{g}|}\frac{\partial\hat{u}}{\partial\tau}
\right)
-\frac{2}{\sqrt{|\hat{g}|}}\frac{\partial}{\partial \tau}
\left(
\hat{g}^{+,-}(s,\tau,y')
\sqrt{|\hat{g}|}\frac{\partial\hat{u}}{\partial s}
\right)
\nonumber
\\
+\sum_{j=1}^{n-1}\frac{2}{\sqrt{|\hat{g}|}}
\frac{\partial}{\partial y_j}
\left(
\hat{g}^{+,j}(s,\tau,y')
\sqrt{|\hat{g}|}\frac{\partial\hat{u}}{\partial s}
\right)     
+\sum_{j=1}^{n-1}
\frac{2}{\sqrt{|\hat{g}|}}\frac{\partial}{\partial s}
\left(
\hat{g}^{+,j}(s,\tau,y')
\sqrt{|\hat{g}|}\frac{\partial\hat{u}}{\partial y_j}  
\right)
\nonumber
\\
+\sum_{j,k=1}^{n-1}\frac{1}{\sqrt{|\hat{g}|}}\frac{\partial}{\partial y_j}
\left(\hat{g}^{j,k}(s,\tau,y')
\sqrt{|\hat{g}|}
\frac{\partial\hat{u}}{\partial y_k}
\right)=0.
\nonumber
\end{eqnarray}
The terms containing $\frac{\partial^2}{\partial s^2},
\frac{\partial^2}{\partial \tau^2},\frac{\partial^2}{\partial y_j\partial\tau}$
vanished because of (\ref{eq:3.1}), (\ref{eq:3.5}).
Here
\begin{equation}                            \label{eq:3.9}
\hat{g}=\left(-4(\hat{g}^{+,-})^{-2}\det[\hat{g}^{jk}]_{j,k=1}^{n-1}\right)^{-1}.
\end{equation} 
It follows from (\ref{eq:3.7}) that
$$
s+\tau-T=\varphi^+(x)+\varphi^-(x),
$$
$$
s-\tau+T=2x_0 +\varphi^+(x)-\varphi^-(x).
$$
Denote (c.f. [E1], (2.23) )
\begin{eqnarray}                          \label{eq:3.10}
y_n=\frac{T-s-\tau}{2}=-\frac{\varphi^+(x)+\varphi^-(x)}{2},
\\
y_0=\frac{s-\tau+T}{2}=x_0+\frac{\varphi^+(x)-\varphi^-(x)}{2},
\nonumber
\\
y_j=\varphi_j(x),\ 1\leq j\leq n-1.
\nonumber
\end{eqnarray}
We shall  also use the coordinates (\ref{eq:3.10}).

Note that $\varphi^+=\varphi^-=0$ when $x_n=0$.   Therefore the map
(\ref{eq:3.10}) is the identity on $x_n=0$.

Since $u_s=\frac{1}{2}(u_{y_0}-u_{y_n}),\ u_\tau=-\frac{1}{2}(u_{y_0}+u_{y_n})$
the equation (\ref{eq:3.8}) has the following form in $(y_0,y',y_n)$
coordinates 
\begin{eqnarray}                         \label{eq:3.11}
\hat{L}\hat{u}=
\hat{g}^{+,-}\frac{\partial^2 \hat{u}}{\partial y_0^2} 
-\frac{1}{\sqrt{|\hat{g}|}}\frac{\partial}{\partial y_n}
\left(\sqrt{|\hat{g}|}\hat{g}^{+,-}(s,\tau,y')
\frac{\partial\hat{u}}{\partial y_n}\right)
\\
+\sum_{j=1}^{n-1}\frac{1}{\sqrt{|\hat{g}|}}
\frac{\partial}{\partial y_j}\left(\sqrt{|\hat{g}|}\hat{g}^{+,j}(s,\tau,y')
\left(\frac{\partial}{\partial y_0}-\frac{\partial}{\partial y_n}\right)\hat{u}\right)
\nonumber
\\
+\sum_{j=1}^{n-1}
\frac{1}{\sqrt{|\hat{g}|}}\left(\frac{\partial}{\partial y_0}-
\frac{\partial}{\partial y_n}\right)\left(\sqrt{|\hat{g}|}
\hat{g}^{+,j}(s,\tau,y')
\frac{\partial\hat{u}}{\partial y_j}\right)
\nonumber
\\
+\sum_{j,k=1}^{n-1}\frac{1}{\sqrt{|\hat{g}|}}\frac{\partial}{\partial y_j} 
\left(\sqrt{|\hat{g}|}\hat{g}^{j,k}(s,\tau,y')
\frac{\partial\hat{u}}{\partial y_k}\right)=0.
\nonumber
\end{eqnarray}
We used above that $\hat{g}^{jk},\hat{g}^+,\hat{g}^{+,j}$  depend on 
$(y',y_n)$  and do not depend on $y_0$.

Divide (\ref{eq:3.11}) by $\hat{g}^{+,-}$.

As in [E1] put
\begin{equation}                    \label{eq:3.12}
u'=|\hat{g}|^{\frac{1}{4}}(\hat{g}^{+,-})^{\frac{1}{2}}\hat{u}.
\end{equation}
Then $u'$ will be the solution of the equation
\begin{eqnarray}                            \label{eq:3.13}
   L_1u'\stackrel{def}{=}u_{y_0^2}'-u_{y_n^2}'
+\sum_{j,k=1}^{n-1}\frac{\partial}{\partial y_j}
\left(g_0^{jk}\frac{\partial u'}{\partial y_k}\right)
\\  
+\sum_{j=1}^{n-1}\left(\frac{\partial}{\partial y_0}-
\frac{\partial}{\partial y_n}\right)\left(g_0^{0j}\frac{\partial u'}{\partial y_j}\right)
\nonumber
\\
+\sum_{j=1}^{n-1}\frac{\partial}{\partial y_j}\left(g_0^{0j}
\left(\frac{\partial}{\partial y_0}-\frac{\partial}{\partial y_n}\right)u'\right)+V_1u'=0,
\nonumber
\end{eqnarray}
where 
$g_0^{jk}=(\hat{g}^{+,-})^{-1}\hat{g}^{jk},\ 
g_0^{0j}=-g_0^{nj}=(\hat{g}^{+,-})^{-1}\hat{g}^{+,j},
\ 1\leq j,k\leq n-1,\ V_1$  has a form similar to (2.8) in
[E1]:
\begin{eqnarray}                               \label{eq:3.14}
\ 
\\
V_1(s,\tau,y')=
\frac{\partial^2A}{\partial y_n^2}+\left(\frac{\partial A}{\partial y_n}\right)^2-
\sum_{j,k=1}^{n-1}\frac{\partial}{\partial y_j}
\left(g_0^{jk}\frac{\partial A}{\partial y_k}\right)
-\sum_{j,k=1}^{n-1} g_0^{jk}\frac{\partial A}{\partial y_j}
\frac{\partial A}{\partial y_k}
\nonumber
\\
+\sum_{j=1}^{n-1}\left(\frac{\partial}{\partial y_n}\left(g_0^{0j}
\frac{\partial A}{\partial y_j}\right) + 
\frac{\partial}{\partial y_j}\left(g_0^{0j}\frac{\partial A}{\partial y_n}
   \right)
+2g_0^{0j}\frac{\partial A}{\partial y_j}\frac{\partial A}{\partial y_n}
\right),
\nonumber
\end{eqnarray}
where
$A=\ln[(\hat{g}^{+,-})^{\frac{1}{2}}|\hat{g}|^{\frac{1}{4}}]=
\ln(\frac{1}{\sqrt{2}}g_1^{\frac{1}{4}}),\ 
g_1=(\det[\hat{g}^{jk}]_{j,k=1}^{n-1})^{-1},\ g_0^{nn}=-1$ (c.f.
(\ref{eq:3.9}) and (\ref{eq:3.12})).

Note that $L_1$is formally self-adjoint.  The DN operator $\Lambda_1$
corresponding to $L_1$ has the following form:
\begin{equation}                                    \label{eq:3.15}
\Lambda_1f'=\left(\frac{\partial u'}{\partial y_n}+
\sum_{j=1}^{n-1}g_0^{nj}\frac{\partial u'}{\partial y_j}\right)|_{y_n=0},
\end{equation}
where $f'=u'|_{y_n=0}$.
It follows from the Remark 2.2 in [E1]  that
$e^A=(\hat{g}^{+,-})^{\frac{1}{2}}|\hat{g}|^{\frac{1}{4}}=
\frac{1}{\sqrt{2}}g_1^{\frac{1}{4}}$ 
and its derivatives on $y_n=0$
can be determined by the DN operator 
$\Lambda$ of $L$.  Therefore the DN operator $\Lambda_1$ of $L_1$  is
determined by the DN operator of $L$ (c.f. [E1],  (2.9)-(2.12)).

Introduce notations similar to [E1],  p. 819.
Let $\Gamma\subset\Gamma^{(1)}\subset\Gamma^{(2)}\subset
U_0\cap\partial\Omega_0$.
Denote by $D_{js_0},1\leq j\leq 2, 0\leq s_0\leq T,$  the forward domain 
of influence of $\overline{\Gamma}^{(j)}\times[s_0,T]$ in the half-space
 $y_n\geq 0$.  
Denote by $D_j^-$  the backward domain of influence  
 of $\overline{\Gamma}^{(j)}\times[0,T]$  for $y_n\geq 0$.

Let $Y_{js_0}, s_0\in[0,T),1\leq j\leq 2,$   be the intersection of 
$D_{js_0}$ with the plane $T-y_n-y_0=0$.  Denote by $X_{js_0}$  the part
of $D_{js_0}$  below $Y_{js_0}$.  Let $Z_{js_0}=\partial X_{js_0}\setminus
(Y_{js_0}\cup\{y_n=0\})$.  We assume that 
$X_{20}\cap\partial\Omega_0\subset U_0$
and $X_{20}$  does not intersect $\partial\Omega$ for $y_n>0$.
We shall call $D_{js_0}\cap D_{j}^-$  the double cone of influence of
$\Gamma^{(j)}\times[s_0,T]$.  Denote by $R_{js_0}$ the intersection of
$\overline{D}_{js_0}\cap\overline{D}_{j}^{\ -}$  with $Y_{js_0}$.

We shall assume that $\Gamma^{(j)},1\leq j\leq 2,$  are such that
$D_{10}\cap\partial\Omega_0\subset\Gamma^{(2)}\times[0,T].$

Let $Q_j$  be the rectangle in the plane $\tau=0:\ Q_j=\{(s,\tau,y'):\tau=0,
0\leq s\leq T,y'\in\overline{\Gamma}^{(j})\}$.  Note that $Q_j$  is the intersection
of $D_j^-$  with the plane $\tau=0$.  Therefore  $R_{js_0}$  is the intersection
of $Y_{js_0}$ with $Q_j,j=1,2.$
Note also that if $(\overline{s},0,\overline{y}')\in Y_{js_0}$
then the line segment $(s,0,\overline{y}'),\overline{s}\leq s\leq t$,
also belongs to $Y_{js_0}$.  
Later we shall introduce one more set $\hat{R}_{10}\subset\overline{\Gamma}^{(1)}
\times[0,T]$  and assume that $\overline{\Gamma}\times [0,T]\subset\hat{R}_{10}$.
We shall refer to this assumption and to the assumptions in the preceeding paragraphs
as the geometric assumptions.  These assumptions can be always satisfied if
$T$  is small.

The following theorem is a generalization of Lemma 2.1  
in [E1]:
\begin{theorem}                        \label{theo:3.1}
Let $\hat{L}^{(1)}$ and $\hat{L}^{(2)}$
be two operators of the form (\ref{eq:3.11})  and let $\hat{\Lambda}^{(i)}$
be the corresponding DN operators.   Assume that 
$\hat{\Lambda}^{(1)}=\hat{\Lambda}^{(2)}$  on $\Gamma^{(2)}\times(0,T)$
and that the  geometric assumptions are satisfied.
Then there exists changes of variables $\hat{y}_0=y_0,\hat{y}_n=y_n,
\hat{y}'=\alpha^{(i)}(y_n,y'),i=1,2,$  such that
$\alpha^{(1)}(0,y')=\alpha^{(2)}(0,y')=y'$ and  
$\tilde{L}^{(1)}=\tilde{L}^{(2)}$  when $\hat{y}'\in\Gamma,\ y_n\in[0,\frac{T}{2}]$.
Here $\tilde{L}^{(i)}$ are differential operators $\hat{L}^{(i)}$  in the 
coordinates $(\hat{y}_0,\hat{y}_n,\hat{y}')$.
\end{theorem}

Many parts of the proof of Theorem \ref{theo:3.1} are the same as in Lemma 2.1
in [E1].  We shall skip the proofs in such cases and concentrate only 
on the new elements.

We shall start with the derivation of Green's formulas analogous to formulas 
(2.33)  and (2.24)  in [E1].

Consider the following initial-boundary value problem for $L_1$:
$$
L_1u=0\ \ \ \ \mbox{for\ } \ y_n>0,
$$
$$
u=u_{y_0}=0\ \ \ \mbox{for\ } \ y_0=0, \ y_n>0,
$$
$$
u|_{y_n=0}=f,
$$
where $\mbox{supp\ } f\subset\overline{\Gamma}^{(2)}\times(0,T],\ 
\Gamma^{(2)}\subset U_0\cap\{y_n=0\}$.

Let $v$ be such that
$$
L_1^*v=0,\ y_n>0,
$$
$$ 
v=v_{y_0}=0 \ \ \  \mbox{when\ }\ y_0=0,\ \ y_n>0,
$$
$$
v|_{y_n=0}=g,\ \ \mbox{supp\ } g\subset\overline{\Gamma}^{(2)}\times(0,T].
$$

We have
$$
0=(L_1u,v)-(u,L_1^*v),
$$
where $(u,v)=\int_{X_{20}}u(y_0,y)\overline{v(y_0,y)}dy_0dy'dy_n$.
Integrating by parts we get
\begin{eqnarray}                            \label{eq:3.16}
(
2\sum_{j=1}^{n-1}\frac{\partial }{\partial s}g_0^{0j}\frac{\partial u}{\partial y_k}
+2\sum_{j=1}^{n-1}\frac{\partial}{\partial y_k}g_0^{0j}\frac{\partial u}{\partial s},v
)
\\
=
(
-2\sum_{j=1}^{n-1}g_0^{0j}\frac{\partial u}{\partial y_k},v_s
)
+
(
-2\sum_{j=1}^{n-1}g_0^{0j}
\frac{\partial u}{\partial s},\frac{\partial v}{\partial y_k}
)
+\int_{y_n=0}\sum_{j=1}^{n-1}g_0^{0j}
\frac{\partial u}{\partial y_k}\overline{v}dy_0dy'
\nonumber
\\
=
(
u,2\sum_{j=1}^{n-1}
\frac{\partial}{\partial y_k}(g_0^{0j}\frac{\partial v}{\partial s})
)
+
(
u,2\sum_{j=1}^{n-1}
\frac{\partial}{\partial s}(g_0^{0j}\frac{\partial v}{\partial y_k})
)
\nonumber
\\
-\int_{y_n=0}u\sum_{j=1}^{n-1}g_0^{0j}\overline{\frac{\partial v}{\partial y_k}}dy_0dy'
+\int_{y_n=0}\sum_{j=1}^{n-1}g_0^{0j}\frac{\partial u}{\partial y_k}\overline{v}dy_0dy'.
\nonumber
\end{eqnarray}
We used here  that
$u,v$  vanish on $Z_{20}$.  Note that other terms in $L_1$  are the same 
as in [E1],  formula (2.33).    Therefore integrating these terms by 
parts as in [E1], (2.33), and combining with (\ref{eq:3.16})  we get the folllowing
Green's formula:
\begin{equation}                                   \label{eq:3.17}
\int_{Y_{20}}\left(\frac{\partial u}{\partial s}\overline{v}
-u\overline{\frac{\partial v}{\partial s}}\right)dsdy'=
-\int_{\Gamma^{(2)}\times[0,T]}(\Lambda_1f\overline{g}-f\overline{\Lambda_1g})dy'dy_0,
\end{equation}
where $\Lambda_1$  is the DN operator (\ref{eq:3.15}).
Note that $L_1^*=L_1$ in our case.
Therefore the left hand side
of (\ref{eq:3.17}) is determined by the boundary data.

Now we shall derive another Green's formula similar to (2.24) in [E1].
Consider
\begin{equation}                  \label{eq:3.18}
0=
(
L_1u,\frac{\partial v}{\partial y_0}
)
+
(
\frac{\partial u}{\partial y_0},L_1v
).
\end{equation}
Integrating by parts in $y_j$ and $s$ we get
\begin{eqnarray}                                      
(
2\sum_{j=1}^{n-1}\frac{\partial}{\partial y_j}g_0^{0j}
\frac{\partial u}{\partial s},v_{y_0}
)
+
(
2\sum_{j=1}^{n-1}\frac{\partial}{\partial s}g_0^{0j}
\frac{\partial u}{\partial y_j},v_{y_0}
)
\nonumber
\\
=(-2\sum_{j=1}^{n-1}g_0^{0j}
\frac{\partial u}{\partial s},v_{y_jy_0}
)
+
(
-2\sum_{j=1}^{n-1}g_0^{0j}
\frac{\partial u}{\partial y_j},v_{y_0s}
)
\nonumber
\\
+\int_{y_n=0}\sum_{j=1}^{n-1}g_0^{0j}
\frac{\partial u}{\partial y_j}\overline{v_{y_0}}dy'dy_0.
\nonumber
\end{eqnarray}
Now integrate by parts in $y_0$ and then again in $s$ and $y_j$.
We get
\begin{eqnarray}                          \label{eq:3.19}
(
2\sum_{j=1}^{n-1}\frac{\partial}{\partial y_j}g_0^{0j}
\frac{\partial u}{\partial s},v_{y_0}
)
+
(
2\sum_{j=1}^{n-1}\frac{\partial}{\partial s}g_0^{0j}
\frac{\partial u}{\partial y_j},v_{y_0}
)
\\
=
-\int_{Y_{20}}\sum_{j=1}^{n-1}g_0^{0j}
\frac{\partial u}{\partial s}\overline{\frac{\partial v}{\partial y_j}}dsdy'
-
\int_{Y_{20}}\sum_{j=1}^{n-1}g_0^{0j}
\frac{\partial u}{\partial y_j}\overline{\frac{\partial v}{\partial s}}dsdy'
\nonumber
\\
+
\int_{y_n=0}\sum_{j=1}^{n-1}g_0^{0j}
\frac{\partial u}{\partial y_j}\overline{\frac{\partial v}{\partial y_0}}dy'dy_0
+\int_{y_n=0}\sum_{j=1}^{n-1}g_0^{0j}
\frac{\partial u}{\partial y_0}\overline{\frac{\partial v}{\partial y_j}}dy'dy_0
\nonumber
\\
-
(
u_{y_0},2\sum_{j=1}^{n-1}
\frac{\partial}{\partial s}(g_0^{0j}\frac{\partial v}{\partial y_j})
)
-
(
u_{y_0},2\sum_{j=1}^{n-1}
\frac{\partial}{\partial y_j}(g_0^{0j}\frac{\partial v}{\partial s})
)
\nonumber
\end{eqnarray}
The remaining terms in (\ref{eq:3.18}) are the same as in [E1],  formulas
(2.18)-(2.25).

Therefore, combining all terms after the integration by parts we get 
(c.f. [E1],  (2.25)):
\begin{eqnarray}                             \label{eq:3.20}
0=(L_1u,v_{y_0})+(u_{y_0},L_1v)
\\
=\tilde{Q}(u,v)+\tilde{\Lambda}_0(f,g),
\nonumber
\end{eqnarray}
where 
\begin{eqnarray}                             \label{eq:3.21}
\tilde{Q}(u,v)=
\ \ \ \ \ \ \ \ \ \ \ 
\ \ \ \ \ \ \ \ \
 \ \ \ \ \ \ \ \ \ \ \
\\
\frac{1}{2}\int_{Y_{20}}\left[4u_s\overline{v_s}
-\sum_{j,k=1}^{n-1}g_0^{jk}
\frac{\partial u}{\partial y_j}\overline{\frac{\partial v}{\partial y_k}}\right.
\left. -2\sum_{j=1}^{n-1}\left(g_0^{0j}\frac{\partial u}{\partial s}
\overline{\frac{\partial v}{\partial y_j}}+
g_0^{0j}\frac{\partial u}{\partial y_j}
\overline{\frac{\partial v}{\partial s}}\right)+V_1u\overline{v}\right]dy'ds
\nonumber
\end{eqnarray}
and 
\begin{equation}                             \label{eq:3.22}
\tilde{\Lambda}_0(f,g)=\int_{\Gamma^{(2)}\times[0,T]}
(\Lambda_1f\overline{g_{y_0}}+f_{y_0}\overline{\Lambda_1g})dy'dy_0.
\end{equation}
Again in the derivation of (\ref{eq:3.20}) we used that $u=v=0$ on $Z_{20}$.

We shall show  now  that the "ellipticity"  condition (\ref{eq:2.3}),
i.e.  that the reduced quadratic form is  negative definite,
implies that $\tilde{Q}(u,v)$ is positive definite.
Note that the map of the form (\ref{eq:3.7}) and, consequently,  the map
(\ref{eq:3.10}),  preserves the ellipticity condition.                    

The reduced quadratic form in (\ref{eq:3.13})
has the form:
\begin{equation}                         \label{eq:3.23}
\sum_{j,k=1}^{n-1}g_0^{jk}(x)\xi_j\xi_k-\xi_n^2
-2\sum_{j=1}^{n-1}g_0^{0j}\xi_j\xi_n.
\end{equation}
 The "ellipticity" 
condition (\ref{eq:2.3}) implies that (\ref{eq:3.23}) is negative definite.
Replacing in the complexification of (\ref{eq:3.23}) $\xi_n$  by $2u_s$  and $\xi_j$  
by $-u_{y_j},\ 1\leq j \leq n-1$,  we get that 
$\tilde{Q}(u,u)$  is positive definite assuming that 
$T$ is small.

Having Green's formulas (\ref{eq:3.20})  with positive
definite $\tilde{Q}(u,u)$  
we can proceed  as in [E1].

Let $L^{(i)}, i=1,2,$  be two operators of the form (\ref{eq:2.1})
and let $(y_0,y)=\Phi_i(x_0,x),i=1,2,$ be two maps
of the form (\ref{eq:3.10})  that transform $L^{(i)}$ to $\hat{L}^{(i)}$
of the form (\ref{eq:3.11}),  $i=1,2.$
Let  $L_1^{(1)}$  and $L_1^{(2)}$  be two operators of the form (\ref{eq:3.13}).

Let $v_i^g,u_i^f,\ i=1,2,$ be such that $L_1^{(i)}u_i^f=0,L_1^{(i)} v_i^g=0$
in $X_{20}^{(i)}, \ u_i^f|_{y_n=0}=f,v_i^g|_{y_n=0}=g,\ u_i^f=v_i^g
=u_{iy_0}^f=v_{iy_0}^g=0$  for
$y_0=0,y_n>0,\ i=1,2$.
We shall denote by 
$[g_{i0}^{jk}]_{j,k=0}^n$  the matrices of $L_1^{(i)}$  in $(y_0,y',y_n)$ 
coordinates,  $i=1,2$.  As in (\ref{eq:3.13}) we have $g_{i0}^{0j}=-g_{i0}^{nj},\ g_{i0}^{+,j}=g_{i0}^{0j}, 
\ g_{i0}^{-,j}=0.$ 

 We assume that $\mbox{supp\ }f$  and $\mbox{supp\ }g$
are contained in $\Gamma^{(2)}\times(0,T]$  and $\Lambda_1^{(1)}=
\Lambda_1^{(2)}$ 
on $\Gamma^{(2)}\times(0,T)$  where $\Lambda_1^{(i)}$ are the DN operators
for $L_1^{(i)}, i=1,2$.   

Let $\Gamma_i^{(j)},D_{js_0}^{(i)},Y_{js_0}^{(i)},X_{js_0}^{(i)},\ j=1,2,$
correspond to $L_1^{(i)},\ i=1,2.$
It was proven in Lemma 2.4 in [E1] that if 
$\Gamma_1^{(1)}=\Gamma_2^{(1)}$
then
$D_{10}^{(1)}\cap\{y_n=0\}=D_{10}^{(2)}\cap\{y_n=0\}$.
Therefore we can take $\Gamma_1^{(2)}=\Gamma_2^{(2)}$,  i.e.
the sets $\Gamma^{(1)},\Gamma^{(2)}$  can be chosen the same for $i=1,2$.

Denote by 
$\stackrel{\circ}{H^1}(Y_{js_0}^{(i)})$ the closure
of $C_0^\infty(Y_{js_0}^{(i)})$  in the Sobolev norm $\|u\|_{1,Y_{js_0}^{(i)}}$
and denote by $H_0^1(Y_{js_0}^{(i)})$  the closure of $C^\infty$ functions in
$Y_{js_0}^{(i)}$ equal to zero on $\partial Y_{js_0}^{(i)}\setminus\{y_n=0\}$.
Analogously one defines $\stackrel{\circ}{H}^1(\Gamma^{j}\times[s_0,T])$ 
and $H_0^1(\Gamma^{(j)}\times[s_0,T])$  (c.f.   [E1]).
\begin{lemma}{(c.f. Lemma 3.4  in [E1]) }                       \label{lma:3.1}
Assuming that $\Lambda_1^{(1)}=\Lambda_1^{(2)}$  on $\Gamma^{(2)}\times(0,T)$
we have
\begin{equation}                              \label{eq:3.24}
C_1\|u_1^f\|_{1,Y_{2s_0}^{(1)}}\leq\|u_2^f\|_{1,Y_{2s_0}^{(2)}}
\leq C_2\|u_1^f\|_{1,Y_{2s_0}^{(1)}}
\end{equation}
for all $f\in H_0^1(\Gamma^{(2)}\times(s_0,T))$.
\end{lemma}

{\bf Proof:}
Applying
the Green's formula (\ref{eq:3.20})  for $i=1,2$  and taking into account that 
$\Lambda_1^{(1)}=\Lambda_2^{(2)}$  we get
$$ 
Q^{(1)}(u_1^f,u_1^f)=Q^{(2)}(u_2^f,u_2^f),
$$
 where  $Q^{(i)}$  corresponds to $L_1^{(i)},i=1,2$.   The inequality
(\ref{eq:3.24})  follows from the ellipticity of $Q^{(i)},i=1,2.$
\qed

Denote by $\Delta_1$  the domain in $\R^{n+1}$ bounded by the planes:
$\Gamma_2=\{\tau=T-y_n-y_0=0,0\leq y_n\leq\frac{T}{2},y'\in\R^{n-1}\},\ 
\Gamma_3=\{s=y_0-y_n=0,\frac{T}{2}\leq  y_n\leq T,y'\in\R^{n-1}\}$
and $\Gamma_4=\{y_0=T,0\leq y_n\leq T,y'\in \R^{n-1}\}$.
Let $L_1$  be an operator of the form (\ref{eq:3.13}) in $\Delta_1$.

\begin{lemma}{(c.f. Lemma 3.1 in [E1]  and Lemma 3.1  in [E3])}
                                                 \label{lma:3.2}
For any $v_0\in \stackrel{\circ}{H^1}(\Gamma_2)$  there exists 
$u\in H^1(\Delta_1),w_0\in \stackrel{\circ}{H^1}(\Gamma_4),w_1\in L_2(\Gamma_4)$
such that $L_1u = 0$  in $\Delta_1,\ u|_{\Gamma_2}=v_0,\ 
u|_{\Gamma_3}=0,\ u|_{\Gamma_4}=w_0,\ u_{y_0}|_{\Gamma_4}=w_1.$
\end{lemma}
{\bf Proof:}

Integrating by part as in the proof  of (\ref{eq:3.20})
and taking into account that $u|_{\Gamma_3}=0$  we get an identity
(c.f. (\ref{eq:3.1})  in [E1]):
\begin{equation}                         \label{eq:3.25}
Q(v_0,v_0)=E(u,u),
\end{equation}
where 
$$
E(u,u)=\int_{\Gamma_4}(|u_{y_0}|^2-\sum_{j,k=1}^ng_0^{jk}u_{y_j}\overline{u}_{u_k}+
V_1|u|^2)dy.
$$
Once the identity (\ref{eq:3.25}) is established, the proof of Lemma \ref{lma:3.2}
 peoceeds as in [E1],  Lemma 3.1.

\begin{lemma}[Density lemma] {(c.f. Lemma 2.2 in [E1])}          \label{lma:3.3}
For any $w\in H_0^1(R_{js_0})$  there exists a sequence $u^{f_n}\in
H_0^1(Y_{js_0}),f_n\in H_0^1(\Gamma^{(j)}\times(s_0,T))$, such that
$$
\|w-u^{f_n}\|_{1,Y_{js_0}}\rightarrow 0\ \ \ \mbox{when}\ \ n\rightarrow\infty.
$$
Note that $H_0^1(R_{js_0})\subset H_0^1(Y_{js_0})$.
\end{lemma}
The proof of Lemma \ref{lma:3.3}  is based on the Green's formula  (\ref{eq:3.20}),
Lemma \ref{lma:3.2}  and the unique continuation theorem of Tataru (c.f. [T]) and it 
is identical to the proof of Lemma 2.2  in [E1].
\qed
  
The main lemma used in the proof of Theorem \ref{theo:3.1} is the following
\begin{lemma}{(c.f. (2.40) in [E1])}       \label{lma:3.4} 
Let $u_i^f,v_i^g, i=1,2,$  be
the solutions of $L_1^{(i)}u=0$  in  $X_{10}^{(i)},i=1,2,$  with
zero initial conditions and $u_i^f|_{y_n=0}=f, \ v_i^g|_{y_n=0}=g$,
where $f,g$  belong to $H_0^1(\Gamma^{(1)}\times[0,T])$.
Suppose   $\Lambda_1^{(1)}=\Lambda_1^{(2)}$  on  $\Gamma^{(2)}\times(0,T)$.  
Then for any $s_0\in [0,T]$  we have
\begin{equation}                               \label{eq:3.26}
\int_{Y_{10}^{(1)}\cap\{s\geq s_0\}}\frac{\partial u_1^f}{\partial s}
\overline{v_1^g}dsdy'=
\int_{Y_{10}^{(2)}\cap\{s\geq s_0\}}\frac{\partial u_2^f}{\partial s}
\overline{v_2^g}dsdy'.
\end{equation}
\end{lemma}

The proof  of Lemma \ref{lma:3.4}  uses 
Lemmas \ref{lma:3.1} and \ref{lma:3.3},  and is exactly the 
same as the proof of (2.40) in [E1].   We shall repeat  this proof here
for the convenience of the reader.

Integrating by parts we obtain
\begin{eqnarray}                       \label{eq:3.27}
\int_{Y_{20}}(u_s^f\overline{v}^g-u^f\overline{v}_s^g)dsdy'=
2\int_{Y_{20}}u_s^f\overline{v}^gdsdy'
\\
\nonumber
-\int_{\Gamma^{(2)}}u^f(T,y',0)\overline{v^g(T,y',0)}dy'.
\end{eqnarray}
Since $u^f(T,y',0)=f(T,y'),\ v^g(T,y',0)=g(T,y')$  we get,
using (\ref{eq:3.17}),  that
$(u_s^f,v^g)\stackrel{def}{=}\int_{Y_{20}}u_s^f\overline{v}^gdsdy'$
is determined by the DN operator.  Therefore we have
\begin{equation}                       \label{eq:3.28}
(\frac{\partial u_1^f}{\partial s},v_1^g)=(\frac{\partial u_2^f}{\partial s},v_2^g)
\end{equation}
for  all $f,g\in H_0^1(\Gamma^{(2)}\times(0,T))$.   Consider
$f,g\in H_0^1(\Gamma^{(1)}\times (0,T))$.  Then 
$\mbox{supp\ }u_i^f$  and $\mbox{supp\ }v_i^g$ 
are contained in $Y_{10}^{(i)},i=1,2.$    Take any  $s_0\in [0,T)$.
It follows from   the geometric 
assumptions that $Y_{10}^{(i)}\cap\{s\geq s_0\}\subset R_{2s_0}^{(i)}$. 
Let $w_i$  be such that $\frac{\partial w_i}{\partial s}=0$   when 
$s>s_0$  and $w_i|_{s=s_0}=u_i^f|_{s=s_0}$.

Let $u_0^{(i)}=u_i^f-w_i$  when  $s\geq s_0,\ u_0^{(i)}=0$  
when $s<s_0$.
Assume that $f$  and therefore $u_i^f$  is smooth.   Then
$u_0^{(i)}\in H_0^1(R_{2s_0}^{(i)})\subset H_0^1(Y_{2s_0}^{(i)})$.     We shall
prove that 
\begin{equation}                               \label{eq:3.29}
(u_{0s}^{(1)},v_1^g)=(u_{0s}^{(2)},v_2^g)
\end{equation}
for any $g\in H_0^1(\Gamma^{(1)}\times [0,T])$.

By Lemma \ref{lma:3.3}  there exists $u_1^{f_n}\in H_0^1(Y_{2s_0}^{(1)})$
such that $\|u_0^{(1)}-u_1^{f_n}\|_{1,Y_{2s_0}^{(1)}}\rightarrow 0$.
By Lemma \ref{lma:3.1} $\|u_2^{f_n}-v^{(2)}\|_{1,Y_{2s_0}^{(2)}}\rightarrow 0$ 
for some $v^{(2)}\in H_0^1(Y_{2s_0}^{(2)})$.   Substituting $f=f_n$  in 
(\ref{eq:3.28}) and passing to the limit when $n\rightarrow\infty$  we get
\begin{equation}                              \label{eq:3.30}
(u_{0s}^{(1)},v_1^g)=(v_s^{(2)},v_2^g).
\end{equation}
Note that (\ref{eq:3.30}) holds for any  $g\in H_0^1(\Gamma^{(2)}\times (0,T))$.
Take $g'\in H_0^1(\Gamma^{(2)}\times (s_0,T))$,  i.e. 
$v_i^{g'}\in H_0^1(Y_{2s_0}^{(i)})$.  Since  
$u_{0s}^{(i)}=\frac{\partial u_i^f}{\partial s}$  when $s\geq s_0$,  and 
$v_i^{g'}=0$  for $s<s_0,i=1,2$  we have  (c.f. (\ref{eq:3.28})
\begin{equation}                                    \label{eq:3.31}
(u_{0s}^{(1)},v_1^{g'})=(u_{0s}^{(2)},v_2^{g'}),\ \ \ 
\forall v_i^{g'}\in H_0^1(Y_{2s_0}^{(i)}).
\end{equation}
Compairing (\ref{eq:3.30})  and  (\ref{eq:3.31})   for $g=g'$,
we obtain  
\begin{equation}                                    \label{eq:3.32}
(u_{0s}^{(2)},v_2^{g'})=(v_s^{(2)},v_2^{g'}).
\end{equation}
Since $v_i^{g'}\in H_0^1(Y_{2s_0}^{(i)})$  is arbitrary  we get by
the Lemma \ref{lma:3.3}  that
\begin{equation}                                 \label{eq:3.33}
v_s^{(2)}=u_{0s}^{(2)}\ \ \ \ \mbox{on}\ \ R_{2s_0}^{(2)}.
\end{equation}
When $g\in H_0^1(\Gamma^{(1)}\times [0,T])$  we have that 
$(\mbox{supp\ }v_2^g)\cap\{s\geq s_0\}\subset Y_{10}^{(2)}\cap\{s\geq s_0\}
\subset R_{2s_0}^{(2)}$.  Therefore we can replace $v_s^{(2)}$  by
$u_{0s}^{(2)}$  in (\ref{eq:3.30}) when $v_2^g\in H_0^1(Y_{10}^{(2)})$ 
and this proves (\ref{eq:3.29}).    Finally,  substracting  (\ref{eq:3.29})
from (\ref{eq:3.28})  we get  (\ref{eq:3.26}).
\qed

The next step of the proof of Theorem \ref{theo:3.1}
will use the geometric optics  solutions.  Since the constructions here 
differ from [E1],  page 824,  we shall proceed with more details.
As in (2.41)  in  [E1]  we are looking for $u_i^f$  in the form:
\begin{equation}                                \label{eq:3.34}
u_i^f=e^{ik(s-s_0)}\sum_{p=0}^N
\frac{1}{(ik)^p}
a_p^{(i)}(s,\tau,y')+u_i^{(N+1)},
\end{equation}
where $k$ is a large parameter, $i=1,2$,
\begin{eqnarray}                             \label{eq:3.35}
4\frac{\partial a_0^{(i)}}{\partial \tau}
-4\sum_{j=1}^{n-1}g_{i0}^{0j}(y)\frac{\partial a_0^{(i)}}{\partial y_j}=0,
\\ 
a_0^{(i)}|_{y_n=0}=\chi_1(s)\chi_2(y'),\ \ \ \ \ i=1,2,
\nonumber
\end{eqnarray}
$a_p^{(i)},p\geq 1$,  satisfy nonhomogeneous equations of the form
(\ref{eq:3.35})  that we will not write 
here and $u^{(N+1)}$ is the same as in (2.41) in [E1]
(c.f. [E1],  page 824).
Here $\chi_1(s)\in C_0^\infty(\R^1), \ \chi_1(s)=1$  for 
$|s-s_0|<\delta,\ \chi_1(s)=0$  for  $|s-s_0|>2\delta,\ \delta$
is small,  $\chi_2(y')\in
C_0^\infty(\Gamma^{(1)})$ is  arbitrary.

Let $\beta_j^{(i)}(y_n,\alpha)$  be the solution of the system
of differential equations
\begin{equation}                                 \label{eq:3.36}
\frac{d\beta_j^{(i)}}{dy_n}=2g_{i0}^{0j}(\beta^{(i)},y_n),\ \ \ 
\beta_j^{(i)}(0,\alpha)=\alpha_j,\ \  1\leq j\leq n-1,\ i=1,2.
\end{equation}
Let $\alpha^{(i)}=\{\alpha_j^{(i)}(y_n,y')\}$ be the inverse to
$\beta^{(i)}=\{\beta_j^{(i)}(y_n,\alpha)\}$.
We have
\begin{equation}                             \label{eq:3.37}
\frac{\partial\alpha_j^{(i)}(\frac{T-s-\tau}{2},y')}{\partial\tau}
-\sum_{k=1}^{n-1}g_{i0}^{k0}(y)\frac{\partial\alpha^{(i)}}{\partial y_k}=0,\ \ 
\alpha_j^{(j)}|_{y_n=0}=y_j,\ \ 1\leq j\leq n-1.
\end{equation}
Therefore 
$a_0^{(i)}(s,\tau,y')=\chi_1(s)\chi_2(\alpha^{(i)}(\frac{T-s-\tau}{2},y'))$ is
the solution of  (\ref{eq:3.35}),  $a_0^{(i)}|_{y_n=0}=\chi_1(s)\chi_2(y')$.
Substituting the geometric optics solutions (\ref{eq:3.34}) in (\ref{eq:3.26}),
integrating by parts and taking the limit when $k\rightarrow\infty$  we obtain
(c.f. (2.42)  in [E1]):
\begin{equation}                              \label{eq:3.38}
\int_{\R^{n-1}}\chi_2(\alpha^{(1)}(\frac{T-s_0}{2},y')\bar{v}_1^g(s_0,0,y')dy'=
\int_{\R^{n-1}}\chi_2(\alpha^{(2)}(\frac{T-s_0}{2},y')\bar{v}_2^g(s_0,0,y')dy'.
\end{equation}
Note that $\tau=0$  on $Y_{10}^{(i)},i=1,2$.
Changing $T$  to $T-\tau',0<\tau'\leq T$  we get (\ref{eq:3.38})
for any $0<\tau<T$.
Consider the following change of coordinates
\begin{equation}                             \label{eq:3.39}
\hat{s}=s,\ \ \hat{\tau}=\tau,\ \ \hat{y}_i=\alpha^{(i)}(\frac{T-s-\tau}{2},y'),
\ \ i=1,2.
\end{equation}
The inverse change of variables has the form:
\begin{equation}                           \label{eq:3.40}
s=\hat{s},\ \ \tau=\hat{\tau},\ \ y'=\beta^{(i)}(\frac{T-\hat{s}-
\hat{\tau}}{2},\hat{y}'),
\ \ i=1,2.
\end{equation}
Note that $y'=\beta^{(i)}(y_n,\hat{y}')$  is the  endpoint of the curve
(\ref{eq:3.36})  starting at $\hat{y}'\in\overline{\Gamma}^{(1)}$
when $y_n=0$  and $\hat{y}'=\alpha^{(i)}(y_n,y')$.

Let $\Sigma=\{(s,\tau): s\geq 0,\tau\geq 0, s+\tau\leq T\}$.
Denote  by $\beta^{(i)}(\Sigma\times\overline{\Gamma}^{(1)})$  the image of
$\Sigma\times\overline{\Gamma}^{(1)}$  under the map (\ref{eq:3.40}), $i=1,2$.
Note that $\beta^{(i)}(\Sigma\times\overline{\Gamma}^{(1)})$ 
is contained in $\overline{X}_{10}^{(i)}$.  Therefore 
$\tilde{R}_{10}^{(i)}\stackrel{def}{=}Q_1\cap\beta^{(i)}(\Sigma\times 
\overline{\Gamma}^{(1)})$  is contained in $R_{10}^{(i)},i=1,2$.
Here $Q_1$  is the rectangle  $\{(s,\tau,y'):\tau=0,s\in [0,T],
y'\in\overline{\Gamma}^{(1)}\}$.
Denote by $\hat{R}_{10}^{(i)}$  the image of $\tilde{R}_{10}^{(i)}$
under the map (\ref{eq:3.39}).
Finally,   denote by $\hat{B}^{(i)}$  the projection  of
$\hat{R}_{10}^{(i)}$  on the plane $y_0=0$.   Note that $\hat{B}^{(i)}
\subset\overline{\Gamma}^{(1)}\times[0,\frac{T}{2}],i=1,2.$
We shall assume that $\hat{B}^{(1)}\supset\overline{\Gamma}\times[0,\frac{T}{2}]$.
This assumption always can be satisfied when $T$  is small enough.

Make the change of variables (\ref{eq:3.40}) in (\ref{eq:3.38}).
We get
\begin{eqnarray}                                \label{eq:3.41}
\int_{\Gamma^{(1)}}\chi_2(\hat{y}')\bar{v}_1^g(s,\tau,\beta^{(1)}
(\frac{T-\tau-s}{2},\hat{y}'))
J_1(y_n,\hat{y}')d\hat{y}'
\\
=
\int_{\Gamma^{(1)}}\chi_2(\hat{y}')\bar{v}_2^g(s,\tau,\beta^{(2)}
(\frac{T-\tau-s}{2},\hat{y}'))
J_2(y_n,\hat{y}')d\hat{y}',
\nonumber
\end{eqnarray}
$y_n=\frac{T-s-\tau}{2},\ J_i(\frac{T-s-\tau}{2},\hat{y}')$  is the Jacobian of the 
map (\ref{eq:3.40}).
Since $\chi_2(y')$  is any $C_0^\infty(\Gamma^{(1)})$  function  we get that for
any $\hat{y}'\in \overline{\Gamma}^{(1)}$
\begin{equation}                              \label{eq:3.42}
v_1^g(s,\tau,\beta^{(1)}(\frac{T-\tau-s}{2},\hat{y}'))
J_1(y_n,\hat{y}')=v_2^g(s,\tau,\beta^{(2)}(\frac{T-\tau-s}{2},\hat{y}'))
J_2(y_n,\hat{y}').
\end{equation}
Note that (\ref{eq:3.42}) holds for $(s,\tau,\hat{y}')
\in\Sigma\times\overline{\Gamma}^{(1)}$.

Let $\chi_1(s)$  be the same as before,  and $\chi_3(y')\in C_0^\infty(\Gamma^{(1)})$
be arbitrary.  Construct $v_{i,k}^g$ as geometric optics solution (\ref{eq:3.34})
with $g=\chi_1(s)\chi_3(y')$.

Take $s=s_0$ and $k\rightarrow\infty$.  We get
\begin{equation}                     \label{eq:3.43}
v_{i\infty}^{g}=\chi_1(s_0)\chi_3(\alpha^{(i)}(s_0,\tau,y')),
\end{equation}
where $v_{i,\infty}^{g}=\lim_{k\rightarrow\infty}v_{i,k}^{g}$.
Substituting $v_{i,k}^g$ in (\ref{eq:3.38})
and taking the limit when $k\rightarrow\infty$ we obtain
\begin{equation}                              
\int_{\R^{n-1}}\chi_2(\alpha^{(1)}(s_0,\tau,y')\chi_3(\alpha^{(1)}(s_0,\tau,y')dy'=
\int_{\R^{n-1}}\chi_2(\alpha^{(2)}(s_0,\tau,y')\chi_3(\alpha^{(2)}(s_0,\tau,y')dy'.
\nonumber
\end{equation}
Make the change of variables  (\ref{eq:3.40}).  Since $\chi_2,\chi_3$  are arbitrary 
we get,  as in (\ref{eq:3.42}),  that
  $J_1(y)=J_2(y)$.
Therefore 
\begin{equation}                              \label{eq:3.44}
v_1^g(s,\tau,\beta^{(1)}(\frac{T-\tau-s}{2},\hat{y}'))
=v_2^g(s,\tau,\beta^{(2)}(\frac{T-\tau-s}{2},\hat{y}')),
\end{equation}
where $(s,\tau,\hat{y}')\in \Sigma\times\overline{\Gamma}^{(1)}$.

Let $w_i^g(s,\tau,\hat{y}')=v_i^g(s,\tau,\beta^{(i)}),i=1,2.$  Then
$w_1^g(s,\tau,\hat{y}')=w_2^g(s,\tau,\hat{y}'),\ \forall(s,\tau,\hat{y}')\in
\Sigma\times\overline{\Gamma}^{(1)}$.

Our strategy to complete the proof of Theorem  \ref{theo:3.1} will be
the following:

Making the changes of variables (\ref{eq:3.40}) in $L_1^{(i)}v_i^g=0$
we get $\tilde{L}_1^{(i)}w_i^g=0,i=1,2$.   Using that  $w_1^g=w_2^g$ 
for all $g\in H_0^1(\Gamma^{(1)}\times(0,T))$  and using  the density
lemma \ref{lma:3.4}
we shall prove that the coefficients of $\tilde{L}_1^{(1)}$  and
$\tilde{L}_1^{(2)}$  are equal.
Since the density property holds for $\tau$  fixed we have to take care 
of terms in $\tilde{L}_1^{(i)}$  that contain derivatives in $\tau$.
\qed

Note that integrating by parts as in (\ref{eq:3.27})  we get
\begin{eqnarray} 
\nonumber
\int_{Y_{20}}(u_s^f\overline{v}^g-u^f\overline{v}_s^g)dsdy'
\\
\nonumber
=-2\int_{Y_{20}}u^f\overline{v}_s^gdsdy'+
\int_{\Gamma^{(1)}}u^f(T,y',0)\overline{v}^g(T,y',0)dy'
\end{eqnarray}
Therefore as in (\ref{eq:3.28})  we conclude that
\begin{equation}                            \label{eq:3.45}
(u_1^f,v_{1s}^g)=(u_2^f,v_{2s}^g).
\end{equation}
Using (\ref{eq:3.45})  instead of (\ref{eq:3.28})  we get an equality of the
form (\ref{eq:3.26}) with the roles of $u^f$  and $v^g$  reversed:

\begin{equation}                            \label{eq:3.46}
\int_{Y_{10}^{(1)}\cap\{s\geq s_0\}}u_1^f\frac{\overline{\partial v_1^g}}{\partial s}dy'ds
=\int_{Y_{10}^{(2)}\cap\{s\geq s_0\}}u_2^f\frac{\overline{\partial v_2^g}}{\partial s}dy'ds.
\end{equation}
From (\ref{eq:3.46}) we get,  analogously to (\ref{eq:3.44}),  that
\begin{equation}                            \label{eq:3.47}
\frac{\partial v_1^g(s,\tau,\beta^{(1)})}
{\partial s}
=\frac{\partial v_2^g(s,\tau,\beta^{(2)}(\frac{T-\tau-s}{2},\hat{y}'))}
{\partial s} \ \ \ \mbox{on\ } \Sigma\times\overline{\Gamma}^{(1)}.
\end{equation}
We used here again that $J_1(y_n,\hat{y}')=J_2(y_n,\hat{y}')$ in 
$\Sigma\times\overline{\Gamma}^{(1)}$.
Differentiating $w_i^g(s,\tau,\hat{y}')=v_i^g(s,\tau,\beta^{i})$
 in $s$ and $\hat{y}'$  we get
\begin{equation}                          \label{eq:3.48}
\frac{\partial w_i^g(s,\tau,\hat{y}')}{\partial s}=
\frac{\partial v_i^g(s,\tau,\beta^{(i)})}{\partial s}
+\sum_{k=1}^{n-1}\frac{\partial v_i^g(s,\tau,\beta^{(i)})}{\partial y_k}\beta_{ks}^{(i)},
\end{equation}
\begin{equation}                          \label{eq:3.49}
\frac{\partial w_i^g(s,\tau,\hat{y}')}{\partial \hat{y}_j}=
\sum_{k=1}^{n-1}\frac{\partial v_i^g(s,\tau,\beta^{(i)})}{\partial y_k}
\frac{\partial \beta_{k}^{(i)}}{\partial \hat{y}_j},
\end{equation}
where $\beta^{(i)}=\beta^{(i)}(y_n,\hat{y}'),\ y_n=\frac{T-s-\tau}{2}$.

It follows from (\ref{eq:3.49}) that
\begin{equation}                        \label{eq:3.50}
\frac{\partial v_i^g(s,\tau,\beta^{(i)})}{\partial y_k}
=\sum_{k=1}^{n-1}
\frac{\partial \alpha_j^{(i)}(\frac{T-s-\tau}{2},\beta^{(i)})}{\partial y_k}
\frac{\partial w_i^g(s,\tau,\hat{y}')}{\partial \hat{y}_j}
\end{equation}
where 
$\left[\frac{\partial \alpha_j^{(i)}(y_n,\beta^{(i)})}{\partial y_k}\right]$
is the inverse matrix to 
$\left[\frac{\partial \beta_{k}^{(i)}(y_n,\hat{y}')}{\partial \hat{y}_j}\right]$.

Substituting (\ref{eq:3.50}) into (\ref{eq:3.48}), using (\ref{eq:3.44}), 
(\ref{eq:3.47}), we get
\begin{equation}                      \label{eq:3.51}
\sum_{j,k=1}^{n-1}\frac{\partial \alpha_j^{(1)}(y_n,\beta^{(1)})}{\partial y_k}
\beta_{ks}^{(1)}
\frac{\partial w_1^g(s,\tau,\hat{y}')}{\partial \hat{y}_j}
=\sum_{j,k=1}^{n-1}\frac{\partial \alpha_j^{(2)}(y_n,\beta^{(2)})}{\partial y_k}
\beta_{ks}^{(2)}
\frac{\partial w_1^g(s,\tau,\hat{y}')}{\partial \hat{y}_j},
\end{equation}
where $y_n=\frac{T-s-\tau}{2},\tau=0,(s,\hat{y}')\in \overline{\Gamma}^{(1)}\times
[0,T]$.

Since $\{v_1^g(s,\tau,y'),g\in C_0^\infty(\Gamma^{(1)}\times(0,T)])\}$
are dense in $\stackrel{\circ}{H^1}(R_{10}^{(1)})$ (c.f. Lemma \ref{lma:3.3}),  
we get that 
$\{w_1^g(s,\tau,\hat{y}')\}$  are dense in $\stackrel{\circ}{H^1}(\hat{R}_{10}^{(1)})$,  
where $\hat{R}_{10}^{(1)}$  is the image of $\tilde{R}_{10}^{(1)}\subset
R_{10}^{(1)}$  under the map
(\ref{eq:3.38}).
Therefore we get (c.f. the end of \$ 2 in  [E3])  that
\begin{equation}                      \label{eq:3.52}
\sum_{k=1}^{n-1}\frac{\partial \alpha_j^{(1)}(y_n,\beta^{(1)})}{\partial y_k}
\beta_{ks}^{(1)}(y_n,\hat{y}')
=\sum_{k=1}^{n-1}\frac{\partial \alpha_j^{(2)}(y_n,\beta^{(2)})}{\partial y_k}
\beta_{ks}^{(2)}(y_n,\hat{y}')
\end{equation}
on $\hat{R}_{10}^{(1)}$.   Here 
$\tau=0,y_n=\frac{T-s}{2}$.  Note that 
$\hat{B}^{(1)}$ is
the projection of 
$\hat{R}_{10}^{(1)}$ on the plane $y_0=0$.   Therefore 
(\ref{eq:3.52}) holds on $\hat{B}^{(1)}$  since $\alpha^{(i)}$  and $\beta^{(i)}$
do not depend on $y_0$.  
We have on $\Sigma\times\overline{\Gamma}^{(1)}$ (c.f. (\ref{eq:3.39}), 
(\ref{eq:3.40})):
\begin{equation}                      \label{eq:3.53}
\alpha_j^{(i)}(\frac{T-s-\tau}{2},\beta^{(i)}(\frac{T-s-\tau}{2},\hat{y}'))=\hat{y}_j,
1\leq j\leq n-1,\ i=1,2.
\end{equation}
Differentiating  (\ref{eq:3.53}) in $s$ we get:
\begin{eqnarray}                      \label{eq:3.54}
\alpha_{js}^{(i)}(\frac{T-s-\tau}{2},\beta^{(i)}(\frac{T-s-\tau}{2},\hat{y}'))
\ \ \ \ \ \ \ \ \ \ \ \
\\
+ 
\sum_{k=1}^{n-1}\alpha_{jy_k}^{(i)}(\frac{T-s-\tau}{2},\beta^{(i)})\beta_{ks}^{(i)}
(\frac{T-s-\tau}{2},\hat{y}'))
=0,\ \ \ \ \ \ i=1,2.
\nonumber
\end{eqnarray}

Combining  (\ref{eq:3.54})  and (\ref{eq:3.52})  we get
\begin{equation}                      \label{eq:3.55}
\alpha_{js}^{(1)}(y_n,\beta^{(1)}(y_n,\hat{y}'))=
\alpha_{js}^{(2)}(y_n,\beta^{(2)}(y_n,\hat{y}')),
\ \ \ 1\leq j\leq n-1,\ \ \ (y_n,\hat{y}')\in \hat{B}^{(1)}.
\end{equation}

Consider the equations
$L_1^{(i)}v_i^g=0$  in $X_{10}^{(i)}$.  
It has the following form in $(s,\tau,y')$  coordinates:
\begin{eqnarray}                             \label{eq:3.56}
L_1^{(i)}v_i^g=-4\frac{\partial^2 v_i^g}{\partial s\partial\tau}
+\sum_{j,k=1}^{n-1}\frac{\partial}{\partial y_j}
\left(g_{i0}^{jk}\frac{\partial v_i^g}{\partial y_k}\right)
\\
+
\sum_{j=1}^{n-1}\left(2\frac{\partial}{\partial s}g_{i0}^{+,j}
\frac{\partial v_i^g}{\partial y_j}+
2\frac{\partial}{\partial y_j}g_{i0}^{+,j}\frac{\partial v_i^g}{\partial s}
\right)
+V_1v_i^g=0,
\nonumber
\end{eqnarray}
where $g_{i0}^{+,j}=g_{i0}^{0j}$.  Note that $g_{i0}^{-,j}$,  i.e.  the
coefficient of $\frac{\partial^2 v_i^g}{\partial\tau\partial y_j}$,
is zero.

Making the change of variables 
(\ref{eq:3.39}) we get equations of the form
\begin{eqnarray}                       \label{eq:3.57}
\tilde{L}_1^{(i)}w_i^g
\stackrel{def}{=}-2J_1^{-1}(y_n,\hat{y}')\left(\frac{\partial}{\partial s}J_1
\frac{\partial w_i^g}{\partial \tau}+
\frac{\partial}{\partial \tau}J_1\frac{\partial w_i^g}{\partial s}
\right)
\ \ \ \ \ \ \ \ \ \ 
\\
-\sum_{j=1}^{n-1}2J_1^{-1}\left(\frac{\partial}{\partial\tau}J_1
\alpha_{js}^{(i)}(y_n,\beta^{(i)})\frac{\partial w_i^g}{\partial\hat{y}_j}
+\frac{\partial}{\partial y_j}J_1
\alpha_{js}^{(i)}(y_n,\beta^{(i)})\frac{\partial w_i^g}{\partial\tau}
\right)
\nonumber
\\
+\sum_{j,k=1}^{n-1}J_1^{-1}\frac{\partial}{\partial\hat{y}_j}
\left(J_1\tilde{g}_{i0}^{jk}
\frac{\partial w_i^g}{\partial \hat{y}_k}\right)
+V_1^{(i)}(y_n,\beta^{(i)})w_i^g(s,\tau,\hat{y}')=0,
\ \ \ (s,\tau,\hat{y}')\in \Sigma\times\overline{\Gamma}^{(1)},
\nonumber
\end{eqnarray}
where
\begin{equation}                               \label{eq:3.58}
\tilde{g}_{i0}^{jk}(y_n,\hat{y}')=\sum_{p,r=1}^{n-1}g_{i0}^{pr}(y_n,\beta^{(i)})
\frac{\partial \alpha_j^{(i)}(y_n,\beta^{(i)}(y_n,\hat{y}'))}{\partial y_p}
\frac{\partial \alpha_k^{(i)}(y_n,\beta^{(i)}(y_n,\hat{y}'))}{\partial y_r},
\end{equation}
$ 1\leq j,k\leq n-1.
$
We used in (\ref{eq:3.57}) that  (c.f. (\ref{eq:3.37}))
\begin{eqnarray}                     \label{eq:3.59}
\tilde{g}_{i0}^{+,j}=\sum_{p=1}^{n-1}g_{i0}^{+,p}(y_n,\beta^{(i)})
\frac{\partial \alpha_j^{(i)}(y_n,\beta^{(i)}(y_n,\hat{y}'))}{\partial y_p}
-\frac{\partial \alpha_j^{(i)}}{\partial\tau}=0,
\\
\tilde{g}_{i0}^{-,j}=
-\frac{\partial \alpha_j^{(i)}(y_n,\beta^{(i)}(y_n,\hat{y}'))}{\partial s},
\ \ \ \mbox{since\ } g_{i0}^{-,j}=0,\ \ 1\leq j\leq n-1.
\nonumber
\end{eqnarray}

Since $w_1^g(s,\tau,\hat{y}')=w_2^g(s,\tau,\hat{y}')$  in 
$\Sigma\times\overline{\Gamma}^{(1)}$,
we have in $\hat{B}^{(1)}$:
\begin{eqnarray}                                   \label{eq:3.60}
(\tilde{L}_1^{(1)}-\tilde{L}_1^{(2)})w_1^g=\sum_{j,k=1}^{n-1}
J_1^{-1}\frac{\partial}{\partial\hat{y}_j}
\left(J_1(\tilde{g}_{10}^{jk}-\tilde{g}_{20}^{jk})\frac{\partial w_1^g}{\partial \hat{y}_k}
\right)
\\
+(V_1^{(1)}(y_n,\beta^{(1)})-V_1^{(2)}(y_n,\beta^{(2)}))w_1^g=0.
\nonumber
\end{eqnarray}
We took into account that $J_1(y_n,\hat{y}')=J_2(y_n,\hat{y}')$ 
holds on $\overline{\Gamma}^{(1)}\times[0,\frac{T}{2}]$ and 
$\ \alpha_{js}^{(1)}(y_n,\beta^{(1)}(y_n,\hat{y}'))=
\alpha_{js}^{(2)}(y_n,\beta^{(2)}(y_n,\hat{y}')),\ 1\leq j\leq n-1$,
holds on $\hat{B}^{(1)}$.

Since $\{w_i^g,g\in C_0^\infty(\Gamma^{(1)}\times(0,T))\}$ are dense in 
$\stackrel{\circ}{H^1}(\hat{R}_{10}^{(1)})$,  
we get,  as in [E3]  (see the end of section 2 in
[E3]), that
\begin{equation}                            \label{eq:3.61}
\tilde{g}_{10}^{jk}=\tilde{g}_{20}^{jk},
\ \ \ 
V_1^{(1)}(y_n,\beta^{(1)})=V_1^{(2)}(y_n,\beta^{(2)})
\ \ \ \mbox{in\ }\ \ \hat{R}_{10}^{(1)}.
\end{equation}
Noting that the coefficients in (\ref{eq:3.61})
do not depend on $y_0$  and $\hat{B}^{(1)}$  
is the projection of $\hat{R}_{20}^{(1)}$
on $y_0=0$  we have that (\ref{eq:3.61})  holds in $\hat{B}^{(1)}$.
Therefore we proved that $\tilde{L}_1^{(1)}=\tilde{L}_1^{(2)}$  in 
$\hat{B}^{(1)}$.  Now we shall prove that also  
$\tilde{L}^{(1)}=\tilde{L}^{(2)}$  in 
$\hat{B}^{(1)}$,  where $\tilde{L}^{(i)}$  is the opeators $\hat{L}^{(i)}$
(see  (\ref{eq:3.8}))  
in $(s,\tau,\hat{y}')$  coordinates.  

Operators $\tilde{L}^{(i)}$ have the following form (c.f. (\ref{eq:3.57})):
\begin{eqnarray}                             \label{eq:3.62}
\tilde{L}^{(i)}=-\frac{2}{\sqrt{|\hat{g}_i|}}
\left(\frac{\partial}{\partial s}\hat{g}_i^{+,-}(y_n,\beta^{(i)}(y_n,\hat{y}'))
\frac{\partial}{\partial\tau}
+\frac{\partial}{\partial\tau}\hat{g}_i^{+,-}(y_n,\beta^{(i)})\frac{\partial}{\partial s}\right)
\\
+\sum_{j,k=1}^{n-1}\frac{1}{\sqrt{|\tilde{g_i}|}}
\frac{\partial}{\partial y_j}\sqrt{|\tilde{g_i}|}\tilde{g}_i^{jk}
\frac{\partial}{\partial y_k}
-\sum_{j=1}^{n-1}2\frac{1}{\sqrt{|\tilde{g_i}|}}
\frac{\partial}{\partial \tau}\sqrt{|\tilde{g_i}|}\hat{g}_i^{+,-}(y_n,\beta^{(i)})
\alpha_{js}^{(i)}(y_n,\beta^{(i)})\frac{\partial}{\partial y_k}
\nonumber
\\
-\sum_{j=1}^{n-1}2\frac{1}{\sqrt{|\tilde{g_i}|}}
\frac{\partial}{\partial y_j}\sqrt{|\tilde{g_i}|}\hat{g}_i^{+,-}(y_n,\beta^{(i)})
\alpha_{js}^{(i)}(y_n,\beta^{(i)})\frac{\partial}{\partial \tau},
\nonumber
\end{eqnarray}
where
$\tilde{g}_i^{jk}$  has the form (\ref{eq:3.58}) with 
$g_{i0}^{pr}$  replaced by 
$\hat{g}_i^{pr}(y_n,\beta^{(i)}(y_n,\hat{y}'))$.  Since
$g_{i0}^{pr}=(\hat{g}_i^{+,-})^{-1}\hat{g}_i^{pr}$
we get that
\begin{equation}                           \label{eq:3.63}
\tilde{g}_i^{jk}(y_n,\hat{y}')=
(\hat{g}_i^{+,-}(y_n,\beta^{(i)}))^{-1}
\tilde{g}_{i0}^{jk}(y_n,\hat{y}').
\end{equation}
We used in (\ref{eq:3.62}) that $\hat{g}_i^{-,j}=0$  for $1\leq j\leq n-1,\ i=1,2,$
and that (\ref{eq:3.37})  implies
$$
\sum_{j=1}^{n-1}\hat{g}_i^{+,j}\frac{\partial\alpha_j^{(i)}}{\partial y_j}
-\hat{g}_i^{+,-}\frac{\partial\alpha_j^{(i)}}{\partial\tau}=0,
$$
since $g_{i0}^{0j}=\frac{\hat{g}_i^{+,j}}{\hat{g}^{+,-}}$.

Therefore to prove that $\hat{L}^{(1)}=\hat{L}^{(2)}$  it remains to prove
that
\begin{equation}                                       \label{eq:3.64}
\hat{g}_1^{+,-}(y_n,\beta^{(1)})=
\hat{g}_2^{+,-}(y_n,\beta^{(2)}).
\end{equation}

Making the change of coordinates (\ref{eq:3.39})
in (\ref{eq:3.14}) we get
\begin{equation}                                     \label{eq:3.65}
V_1^{(i)}(y_n,\beta^{(i)}(y_n,\hat{y}'))=
-\sum_{j,k=1}^{n}J_i^{-1}\frac{\partial}{\partial\hat{y}_j}
\left(J_i\tilde{g}_{i0}^{jk}
\frac{\partial \tilde{A}^{(i)}}{\partial \hat{y}_k}\right)
-\sum_{j,k=1}^{n}
\tilde{g}_{i0}^{jk}
\frac{\partial  \tilde{A}^{(i)} }{\partial\hat{y}_j}
\frac{\partial \tilde{A}^{(i)}}{\partial \hat{y}_k},
\end{equation}
where $\hat{y}_n=y_n,\ \tilde{A}^{(i)}(y_n,\hat{y}')=
A^{(i)}(y_n,\beta^{(i)}(y_n,\hat{y}')),\ \tilde{g}_{i0}^{jk},
1\leq j,k\leq n-1$,  are the same as in (\ref{eq:3.57}),
$\tilde{g}_{i0}^{jn}=\tilde{g}_{i0}^{nj}=-\alpha_{js}^{(i)}(y_n,\beta^{(i)}),
1\leq j\leq n-1,\ \tilde{g}_{i0}^{nn}\equiv -1$.

Taking into account that $\tilde{g}_{10}^{jk}=\tilde{g}_{20}^{jk},\ 
J_1=J_2$, 
and 
$$
\tilde{A}_{\hat{y}_j}^{(1)}\tilde{A}_{\hat{y}_k}^{(1)}-
\tilde{A}_{\hat{y}_j}^{(2)}\tilde{A}_{\hat{y}_k}^{(2)}=
(\tilde{A}_{\hat{y}_j}^{(1)}-
\tilde{A}_{\hat{y}_j}^{(2)})\tilde{A}_{\hat{y}_k}^{(1)}+
(\tilde{A}_{\hat{y}_k}^{(1)}-\tilde{A}_{\hat{y}_k}^{(2)})
\tilde{A}_{\hat{y}_j}^{(2)},
$$
 we can rewrite
$$
0=V_1^{(1)}(y_n,\beta^{(1)})-V_1^{(2)}(y_n,\beta^{(2)})
$$
as homogeneous second order elliptic equation for 
$A^{(1)}(y_n,\beta^{(1)})-A^{(2)}(y_n,\beta^{(2)})$,
where $A^{(i)}(y_n,y')=\ln((\hat{g}_i^{+,-}(y))^{\frac{1}{2}}
|\hat{g}_i|^{\frac{1}{4}})=
\ln(\frac{1}{\sqrt{2}}(\det[\hat{g}_i^{jk}(y)]_{j,k=1}^{n-1})^{-\frac{1}{4}}$ 
(c.f. (\ref{eq:3.9})).
Since $\tilde{A}^{(1)}$  and $\tilde{A}^{(2)}$ have the same Cauchy data when
$y_n=0$  (see Remark 2.2 in [E1])  we get,  by the unique continuation theorem 
for the elliptic equations,  that
$A^{(1)}(y_n,\beta^{(1)})=A^{(2)}(y_n,\beta^{(2)})$ in $\hat{B}^{(1)}$.
Since 
$$
\hat{g}_i^{jk}(y_n,\beta^{(i)})=\frac{g_{i0}^{jk}(y_n,\beta^{(i)})}
{\hat{g}_i^{+,-}(y_n,\beta^{(i)})}
$$
and
$$
g_{10}^{jk}(y_n,\beta^{(1)})=g_{20}^{jk}(y_n,\beta^{(2)})
$$
we get (\ref{eq:3.64}).  Therefore
$\tilde{L}^{(1)}=\tilde{L}^{(2)}$ in $\hat{B}^{(1)}$.  Note that 
by the assumption
$\hat{B}^{(1)}\supset\overline{\Gamma}\times[0,\frac{T}{2}]$.
\qed

Theorem \ref{theo:3.1}  concludes the local step of the proof of 
the main Theorem \ref{theo:2.3}.  The global step of the proof is
similar to the proof in [E2]:

Consider the initial-boundary value problems for $L^{(i)}u_i=0$  in
domains $\Omega^{(i)}=\Omega_0\setminus\cup_{j=1}^{m_i}\Omega_j^{(i)},
\ i=1,2$.
Let $\overline{\Delta}_i\subset\Omega^{(i)}$  be the image of 
$\varphi_i^{-1}\circ\alpha^{(i)}(\overline{\Gamma}\times[0,\frac{T}{2}])$,  
where
$\alpha_i$  is the map  (\ref{eq:3.39})  and $\Phi_i(x_0,x)=
(x_0+a_i(x),\varphi_i(x))$  is the map
(\ref{eq:3.10}).  Denote by $\Phi_3$  the map
$\Phi_3=\Phi_1^{-1}\circ\alpha^{(1)}\circ\beta_2\circ\Phi_2$,  where
$\beta_2$  has the form  (\ref{eq:3.40}).  Note that $\Phi_3$  is a
diffeomorphism of the form (\ref{eq:3.10}),  $\ \Phi_3=I$  on 
$(\Delta_2\cap\partial\Omega_0)\times(-\infty,+\infty)$  and
$$
\Phi_3\circ L^{(2)}=L^{(1)}\ \ \ \mbox{on\ } \Delta_1.
$$
Note that  any map $\Phi$  of the form (\ref{eq:3.10})  can be
represented as a composition $\Phi=a_1\circ \varphi_1=\varphi_2\circ a_2$,
where $\varphi_i$  are the diffeomorphisms of $\overline{\Delta}_2$ onto
$\overline{\Delta}_1$
and maps $a_i$ have the form $y_0=x_0+a_i(x),\ y=x,\ a_i(x)\in C^\infty,\ 
a_i(x)=0$  on $\partial\Omega_0$.

It follows from [Hi],  Chapter 8,  that there exists an extension $\tilde{\Phi}_3$
of the map $\Phi_3$ such that $\Phi_3|_{\partial\Omega_0\times(-\infty,\infty)}=I,\ 
\Phi_3$  has a form (\ref{eq:3.10}),  i.e.  $\Phi_3=a_3\circ\varphi_3,\ \varphi_3$
is a diffeomorphism of $\overline{\Omega}^{(2)}$  onto  
$\overline{\Omega}^{(3)}\stackrel{def}{=}\varphi_3(\overline{\Omega}^{(2)})$.
Denote $L^{(3)}=a_3\circ\varphi_3\circ L^{(2)}$.
Then $L^{(3)}$  is a differential operator of the form (\ref{eq:2.1})
on $\Omega^{(3)},\ \Delta_1\subset \Omega^{(3)}$ 
and  $L^{(3)}=L^{(1)}$  on $\Delta_1$.

The proof of the following lemma is the same as in [E1],  Lemma 3.3
(c.f. [KKL1], Lemma 9):
\begin{lemma}                                 \label{lma:3.5}
Let $\Delta_1'\subset\Delta_1$  be such that 
$\Omega_1\setminus\overline{\Delta}_1'$ has 
a smooth boundary,  $\gamma_1=\partial\Omega_0\cap\partial\Delta_1'$  is connected and
$L^{(1)}=L^{(3)}$  on $\Delta_1'$.  Let  $\gamma_2=\partial\Delta_1'
\setminus\gamma_1$.
Suppose $\Lambda^{(1)}=\Lambda^{(2)}$  on $\partial\Omega_0\times(-\infty,+\infty)$,
where $\Lambda^{(i)}$  are DN operators corresponding to 
$L^{(i)}$,  respectively, $i=1,2,3.$   Note that $\Lambda^{(3)}=\Lambda^{(2)}$
on $\partial\Omega_0\times(-\infty,+\infty)$.  Then the DN operators $\Lambda_1^{(1)},
\Lambda_1^{(3)}$ corresponding to the operators $L^{(1)},L^{(3)}$  in the smaller
domains  $\Omega^{(1)}\setminus\overline{\Delta}_1', 
\Omega^{(3)}\setminus\overline{\Delta}_1'$  are equal on 
$((\partial\Omega_0\setminus\overline{\gamma}_1)\cup\gamma_2)\times(-\infty,+\infty)$.
\end{lemma}

Therefore Theorem \ref{theo:3.1} and Lemma \ref{lma:3.5}
reduce the inverse problem for $L^{(1)},L^{(2)}$  in 
$\Omega^{(1)}\times(-\infty,+\infty),\ \Omega^{(2)}\times(-\infty,+\infty)$ to the 
inverse problem for $L^{(1)},L^{(3)}$ in smaller domains
$(\Omega^{(1)}
\setminus\overline{\Delta}_1')\times
(-\infty,+\infty), (\Omega^{(3)}\setminus\overline{\Delta}_1')\times
(-\infty,+\infty)$. 

Continuing this process as in [E2] we can prove the main Theorem \ref{theo:2.3}.
Note that it is enough to have $\Lambda^{(1)}=\Lambda^{(2)}$ on 
$\Omega\times(0,T_0)$,  where  $T_0$  is large enough, to prove
Theorem \ref{theo:2.3}.


\begin{thebibliography}{9999}
\bibitem[AB]{} Aharonov, Y. and Bohm, D., Significance of
electromagnetic potentials in quantum theory,
Phys.Rev., Second Series 115, 485-491 (1959)
\bibitem[B]{} Belishev, M., 1997, Boundary control in reconstruction
of manifolds and metrics (the BC method),
Inverse Problems 13, R1-R45
\bibitem[BCLUW]{} Berry, M., Chambers, R., Large, M., Upstill, C., 
Walmsley, J., 1980, Eur. J. Phys. 1, 154
\bibitem[CFM]{} Cook, R., Fearn, H., Millouni, P., 1995, 
Am. J. Phys. 63, 705
\bibitem[E1]{} Eskin, G., 2006, A new approach to the hyperbolic
inverse problems,  Inverse problems, vol. 22, No. 3
\bibitem[E2]{} Eskin, G., 2007, A new approach to the hyperbolic
inverse problems II, (Global step),  ArXiv:math,AP/07013
(to appear in Inverse Problems)
\bibitem[E3]{} Eskin, G., 2006, Inverse  hyperbolic problems 
with time-dependent coefficients,  ArXiv:math.AP/050816, v.2
(to appear in Comm. in PDE)
\bibitem[E4]{} Eskin, G., 2006, Inverse problems for the Schr\"{o}dinger
equations with time-dependent electromagnetic potentials
and the Aharonov-Bohm effect, ArXiv:math.AP/0611342
\bibitem[E5]{} Eskin, G., 2004, Inverse boundary value problems
in domains with several obstacles,  Inverse problem 20, 1497-1516
\bibitem[ER]{} Eskin, G.,  and Ralston, J., 1997, Inverse scattering problem
for the Schr\"{o}dinger equation with magnetic 
and electric potentials, The IMA Volumes in Mathematics and its applications,
vol 90 (New York: Springer), 147-166
\bibitem[ER1]{} Eskin, G.,  and Ralston, J., 1995, Inverse scattering problem
for the Schr\"{o}dinger equation with magnetic potential at a fixed energy,
Comm. Math. Phys. 173, 199-224
\bibitem[G]{} Gordon, W., 1923, Ann. Phys.  (Leipzig) 72, 421
\bibitem[Hi]{} Hirsch, M., 1976, Differential Topology (New York:Springer)
\bibitem[KKL]{} Katchalov, A., Kurylev, Y., Lassas, M., 2001,
Inverse boundary spectral problems (Boca Baton : Chapman\&Hall)
\bibitem[KKL1]{} Katchalov, A., Kurylev, Y., Lassas, M., 2004,
Energy measurements and equivalence of boundary data for inverse problems
on noncompact manifolds,  IMA Volumes, v.137, 183-214
\bibitem[KL]{} Kurylev, Y. and Lassas, M., 2000,
Hyperbolic inverse problems with data on a part of the boundary, 
AMS/1P Stud. Adv. Math, 16, 259-272
\bibitem[KKLM]{} Katchalov, A., Kurylev, Y., Lassas, M., Mandache, N., 2004,
Equivalence of time-domain inverse problems and boundary spectral problems,
Inverse problems 20, No 2, 419-436
\bibitem[LP]{} Leonhardt, V., Philbin, T., 2006,  General relativity
in Electrical Engineering, New  J.Phys. 8, 247 
\bibitem[LP1]{} Leonhardt, V., Piwnicki, P., 1999,  Phys.  Rev. A60,  4301
\bibitem[LP2]{} Leonhardt, V., Piwnicki, P., 2000,  Phys.  Rev. Lett. 84,  822
\bibitem[LU]{} Lee, J. and Uhlmann, G., 1989, Determining anisotropic 
real-analytic conducivity by boundary measurements,  
Comm. Pure Appl. Math. 42, 1097-1112
\bibitem[N]{} Nicoleau, F., An inverse scattering problem with
the Aharonov-Bohm effect, Journ. Math. Phys., 41, 5223-5237 (2000)
\bibitem[NSU]{} Nakamura, G., Sun, Z., Uhlmann, G.,
Global identificability for inverse problem for the Schr\"{o}dinger
equation in a magnetic field,  Math. Ann. 303, 377-88
\bibitem[NVV]{} Novello, M., Visser, M.,  Volovik, G. (editors),
Artificial black holes, 2002,  World Scientific,  Singapore.
\bibitem[OD]{} O'Dell, S.,  Inverse scattering for the Laplace-Beltrami 
operators  with complex-valued electromagnetic potentials and
embedded obstacles,
Inverse problems 22 No 5 (2006),  1579-1603
\bibitem[OP]{} Olarin, S.  and  I. Iovitzu Popescu, 1985,
The quantum effects of electromagnetic fluxes,  Review of Modern Physics,
vol.  57,  N2,  339-436 
\bibitem[P]{} Pham Mau Quan, 1957,  Archive for Rat. Mech. Anal., 1, 54
\bibitem[RdeRTF]{} Roux, P., de Rosny J., Tanter, M., Fink, M.,
1997,  Phys. Rev. Lett.  79, 317
\bibitem[VMCL]{} Vivanco, F., Melo, F., Coste, C., Lund, F., 1999,
Phys. Rev. Lett. 83, 1966
effect and time-dependent
inverse scattering theory, preprint (2001)
\bibitem[W]{} Weder, R. The Aharonov-Bohm effect and time-dependent
inverse scattering theory, Inverse problems, vol. 18, 1041
\bibitem[WY]{} Wu, T. and Yang, C., Phys. Rev. D 12,3845 (1975)





\end{thebibliography}
\end{document}